\documentclass[%
 reprint,
showpacs,preprintnumbers,
 amsmath,amssymb,
 aps,
prb,
]{revtex4-1}

\usepackage{natbib}

\usepackage{graphicx}
\usepackage{dcolumn}
\usepackage{bm}
\usepackage{mathtools}
\usepackage{hyperref}

\begin{document}


\title{Hysteresis assisted narrowband resonances in a chain of nonlinear plasmonic arrays}


\author{S.~V.~Fedorov$^{1,3}$, A.~V.~Chipouline$^{2,1}$, N.~N.~Rosanov$^{1,3}$}
\affiliation{$^1$National Research University of Information Technologies, Mechanics and Optics (University ITMO), 197101 St.Petersburg, Russia,\\
$^2$Institute of Applied Physics, Friedrich Schiller University, Jena, Germany\\
$^3$Vavilov State Optical Institute, 199034 St.Petersburg, Russia}

\date{\today}

\begin{abstract}
The plasmonic structures exhibiting narrowband resonances (NBR) are of a great interest for various applications. We propose to use hysteresis behavior in a 1D system of nonlinear nanoresonators in order to achieve the NRB; the nonlinearity is provided by saturation of a two-level quantum system coupled with the nanoresonators (nanolaser/spaser configuration). Quantum Dots (\textsl{QD}) were assumed as quantum systems; their numerical parameters have been adopted for estimations. Role of the loss compensation on the quality of the NBR is shown for below (under compensation) and above threshold (generating spasers) operation modes. Amplitude and phase detection schemes of the prospective experimental realization are compared using the developed model. Possible sensor oriented applications of the proposed system are discussed. 

\end{abstract}


\maketitle


\section{\label{sec:Intro}Introduction}

Nanophotonic structures of various natures and shapes receive explosively growing attention during last decade. These structures promise great potential for applications in wide range of science and technology \cite{ZheludevRoadmap11} covering practically all areas of modern life. Among the other, the plasmonic structures are used for various sensor applications; both propagating and localized plasmons are well suited for different particular tasks and promise unprecedented sensitivity and selectivity. Both amplitude and phase response of the nanoplasmonic structures were shown to be useful for the sensing applications 
\cite{Kabashin97,Kabashin98,Kabashin09,Kravetz13}

One of the main drawbacks of plasmonic nanostructures, restricting their potential application, is the intrinsic (ohmic) losses caused by the interaction of the free electrons of the metal with thermo bath (irreversible losses) and radiative losses. The more light is localized to the metal surface, the more plasmonic field fraction is concentrated inside the metal resulting in the appearance of higher dissipative losses \cite{Maier06}. Passive losses as a limiting factor was pointed out a rather long time ago \cite{Boardman82}$^,$\cite{Raether88} and only recently new materials with reduced losses have been suggested 
\cite{Sokoulis10,Boltasseva11,Anlage11,Chen10}. 
Nevertheless, keeping in mind metal as a main candidate for the plasmonic components, the only way to compensate the losses is to use optically active materials in combination with the nanostructures 
\cite{Berini11,Sokoulis11,Barnes98,Ramakrishna03,Bergman03}

The high level of the passive losses stipulates relatively wide bandwidth of the plasmonic resonances, which restricts their sensitivity and selectivity. In this paper we investigate an object, which promises to exhibit ultra-narrow band (ultra-NRB) resonances, namely chains of optically coupled nanolasers/spasers. The chain of the optically coupled nanoresonators was shown to exhibit narrow resonances \cite{Markel05}$^,$\cite{Kravets08}; the nanolaser/spaser also naturally tends to generate narrowband spectrum lines \cite{Noginov09}$^,$\cite{Chirkin11}. Combining these two approaches, it is expected that the chain of the optically coupled nanolasers/spasers will generate the NRB resonances applicable for the sensing applications.  

The model of the 1D chain of optically coupled nanolasers/spasers has been recently theoretically considered \cite{Andrianov12}. In this paper the only interaction between neighboring nanolasers/spasers has been taken into account, which excludes appearance of the NRB resonances; the latter requires interaction not only between next spasers in a chain, but rather interaction of each spaser with all other spasers taking into account retardation (see\cite{Markel05}, where source of the narrowband resonances has been clearly shown for the passive case).  

In order to describe properly the dynamics of the coupled nanolasers, it might be necessary first recall the results from two areas of optical plasmonics, namely: the dynamics of a single nanolaser/spaser, and the dynamics of the coupled nanoplasmonic resonators excited by an external field. Combining these two approaches accepted, a solid background for investigation of the chains of the nanolasers/spasers will be created. 

The nanoresonator changes the radiative properties of the quantum system coupled to them \cite{Purcell46}$^,$\cite{Koenderink10} and can cause both enhancement \cite{Blombergen54}$^,$\cite{Strandberg57} or inhibition \cite{Bunkin59} of spontaneous emission. Nevertheless, in the case of regular dynamics of a nanolaser, spontaneous emission does not affect the dynamics when operating well above threshold; influence of the Purcell effect can be taken into account by appropriate choice of the phenomenological coefficients in the model.                

The nanolaser dynamics is based on energy transfer from the excited quantum emitter to the plasmons, and therefore depends strongly on the positioning of the emitters near the nanoresonator \cite{Hess12}. For example, an appropriate positioning of the emitters can enhance generation of bright and suppress generation of dark modes, and vice versa \cite{Hess12}$^,$\cite{Liu09}. The emitters appear to be coupled with the plasmonic modes from ones side and with the far field (radiative) modes from another one. Moreover, the radiative losses can exceed dissipative ones by a factor of two \cite{Rockstuhl06}$^,$\cite{Husnik08}. In the model developed here, the radiative losses are included in the damping coefficient $\gamma$ for a plasmonic mode. An appropriate positioning of the emitters can redistribute energy transfer in favor of a plasmonic mode which is more effective near sharp angles by a factor of $\left(kr\right)_{}^{3} $, where \textit{k} is the wave vector and \textit{r} is the curvature of the shape of the nanoresontor 
\cite{Chang06,Martin10,Klimov08}
Nevertheless, positioning the emitters too close to the metallic surface can cause quenching of inversion and so must be additionally avoided \cite{Klimov10}.   

The principles of nanolaser design is suggested and developed in 
\cite{Bergman03}$^,$\cite{Stockman08,Stockman10,Stockman11}
and was later experimentally realized in various different configurations \cite{Noginov09}$^,$\cite{Zheludev08,Oulton09,Hill09,Zhu09,Banerjee09,Nezhad10,Ma11,Flynn11,Wu11}.
Recent achievements in this area are summarized in several review articles \cite{Berini11}$^,$\cite{Hess12}$^,$\cite{Yin11}.     

Theoretical models of the nanolaser can be approximately divided by fully numerical \cite{Gordon07}$^,$\cite{Wuestner11} and semi-analytical \cite{Stockman11}$^,$\cite{Sarychev07}$^,$\cite{Andrianov11}, with the model developed here belonging to the latter approach. In both versions, the quantum dynamics of the emitters is described by the density matrix method \cite{Fano57}$^,$\cite{Fain72} adopted for two, three, or four level schemes. The main difference is in the description of the plasmonic oscillations. In order to make the model treatable (at least to some extent) analytically, the plasmon dynamics can be reduced to some version of the harmonic oscillator equation, which finally results in the well-known point-like dynamic laser model \cite{Haken85}. This model has been used many times for investigation of the laser dynamics as a self-oscillating system, and as a modeling task for various problems of nonlinear (including stochastic) dynamics, stability analysis, etc. \cite{Haken85,Akhmanov81,Pikovsky01}. To some extent, this model is especially well suited for the nanolaser due to the monomode (or two-mode) oscillation regime \cite{Protsenko05}, while the usual ``macro'' laser tends to operate in multimode regime. Nevertheless, there are several new tasks, which have not been previously addressed, or have not been addressed fully in a consistent manner. One task, for example, is the generation of a nanolaser operating with bright and dark modes simultaneously (mode competition), and another one is the problem of bandwidth of the nanolaser oscillation. 

It should be emphasized, that the rigorous numerical calculations undoubtedly provide results closer to the experimental ones, but at the same time partially hide the physical insight of the problem. In order to understand necessity of an analytical treatment, consider the problem of instability of the nanolaser operation under the action of an external field. In case of the numerical approach there is no way to subdivide generated by the plasmons and incoming one fields \cite{Wuestner11}, while analytical modeling provides clear qualitative explanation of appearance of the unstable regimes \cite{Andrianov11}.        

The interaction between the small particles (meta-atoms), either dielectric or metallic, and the propagation of an optical excitation in a regular chain of such particles has been extensively investigated 
\cite{Deng10,Weber04,Quinten98,Gippius10,Feth10,Alu06}
Interest in the behavior of chains of metallic nanoparticles was driven mainly by the pursuit of subwavelength guiding structures for a new generation of the optoelectronic components in the area of communication and information processing. Nevertheless, theoretical tools for the modeling of these chains (irrespective to the nature and sizes) remain invariant: ideally the electromagnetic excitation in the particles is supposed to be described by taking into account all possible eigenmodes \cite{Deng10}$^,$\cite{Quinten98} and interactions between all particles in a chain. There are several approximations which are typically accepted in these kinds of problems. Firstly, depending on the size of the particles, the model can be restricted by consideration of dipole moment only (for metallic nanoparticles) \cite{Weber04}$^,$\cite{Alu06}; the higher moments and magnetic response can be also taken into consideration 
\cite{Feth10}$^,$\cite{Rico-Garcia12}. Usually, for the problem of electromagnetic excitation propagation the dipole approximation is enough \cite{Maier03}, provided distance between particles is not less than about three times their dimensions. Secondly, the interaction between the particles in the frame of the quasi-static approximation assumes no retardation; otherwise interaction between dipoles contains terms proportional to the $1/r$ and $1/r^{2}$ in addition to the quasistatic term of  $1/r^{3}$ ($r$ is the distance between dipoles). The problem possesses an exact solution for the infinite chain in the quasistatic limit, while taking into consideration the retardation leads to known mathematical difficulties and requires continuation into the lower half frequency plane \cite{Weber04}. Consideration of the finite chain is free from these excessive mathematical problems, but can be treated only numerically; the respective solutions for both longitudinal and transverse modes are presented in \cite{Weber04}$^,$\cite{Alu10}.  

In this paper, a comprehensive theoretical study of the dynamics of the 1D chain of the spasers/nanolasers is considered. In the model, each nanolaser in the chain interacts with all other nanolasers in analogy with the model of \cite{Markel05} taking into account retardation between the elements. The well-known effect of the appearance of diverged sums in the unlimited chain of coupled resonators has been extended on the case of active elements, namely nanoresonators coupled with the optically active molecules (dye, quantum dots). Both operation modes -- partial loss compensation (below lasing threshold) and complete loss compensation (above lasing threshold) -- have been investigated.    

The paper consists of four parts including this introduction. In the next (second) part, a model to describe the response of the periodically placed nano particles (\textsl{NP}) coupled with the quantum dots (\textsl{QD}) is elaborated. In the third part, the model is applied for the 1D homogeneous periodic chain (i.e. all pairs \textsl{NP}-\textsl{QD} are assumed to be identical). Results and conclusions summarize the paper in the fifth part.      

\section{\label{sec:Mod}The model}

It is believed, that the semiclassical model \cite{Chipouline12} is fully appropriate for the considered in the paper system. The quantum dynamics of the active molecules (dye, quantum dots) is described by the well established in laser physics density matrix formalism, while electron dynamics in plasmonic nanoresonators is fully classical and could be to the first approximation modelled by harmonic oscillator equations \cite{Petschulat08}. In contrast with the other publications \cite{Stockman10}$^,$\cite{Andrianov12}, an electric field and respective plasmonic modes are described classically from the beginning, as like as the density matrix operators are used in basis of eigen functions in energy representation. It is stipulated by the fact, that at the stabilized operation the number of photons is huge and there is no necessity to use the quantum optics tools of secondary quantization (birth and annihilation operators). An initial stage of the generation (which starts from the spontaneous emission) undoubtedly requires quantum optics for an adequate consideration; from the other side, in all referred here papers this transition stage has not been considered and the equations were finally used to describe a steady state operation. 

\begin{figure}[b]
\includegraphics[width=0.4\textwidth]{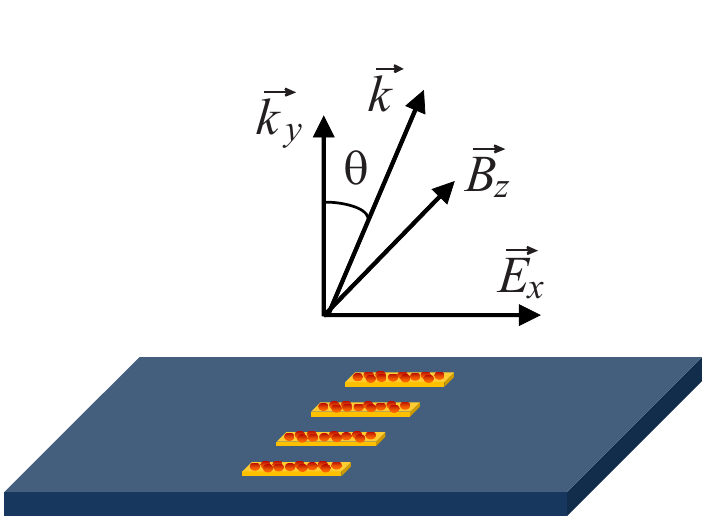} 
\caption{\label{fig1} Possible experimental realization of the 1D chain of the 
\textsl{NP}-\textsl{QD}. The yellow bricks represent plasmonic nanoresonators (nanoparticles -- \textsl{NP}), while active molecules (i.e. Quantum Dots - \textsl{QD}) are shown by small red circles.}
\end{figure}

\begin{figure}[b]
\includegraphics[width=0.4\textwidth]{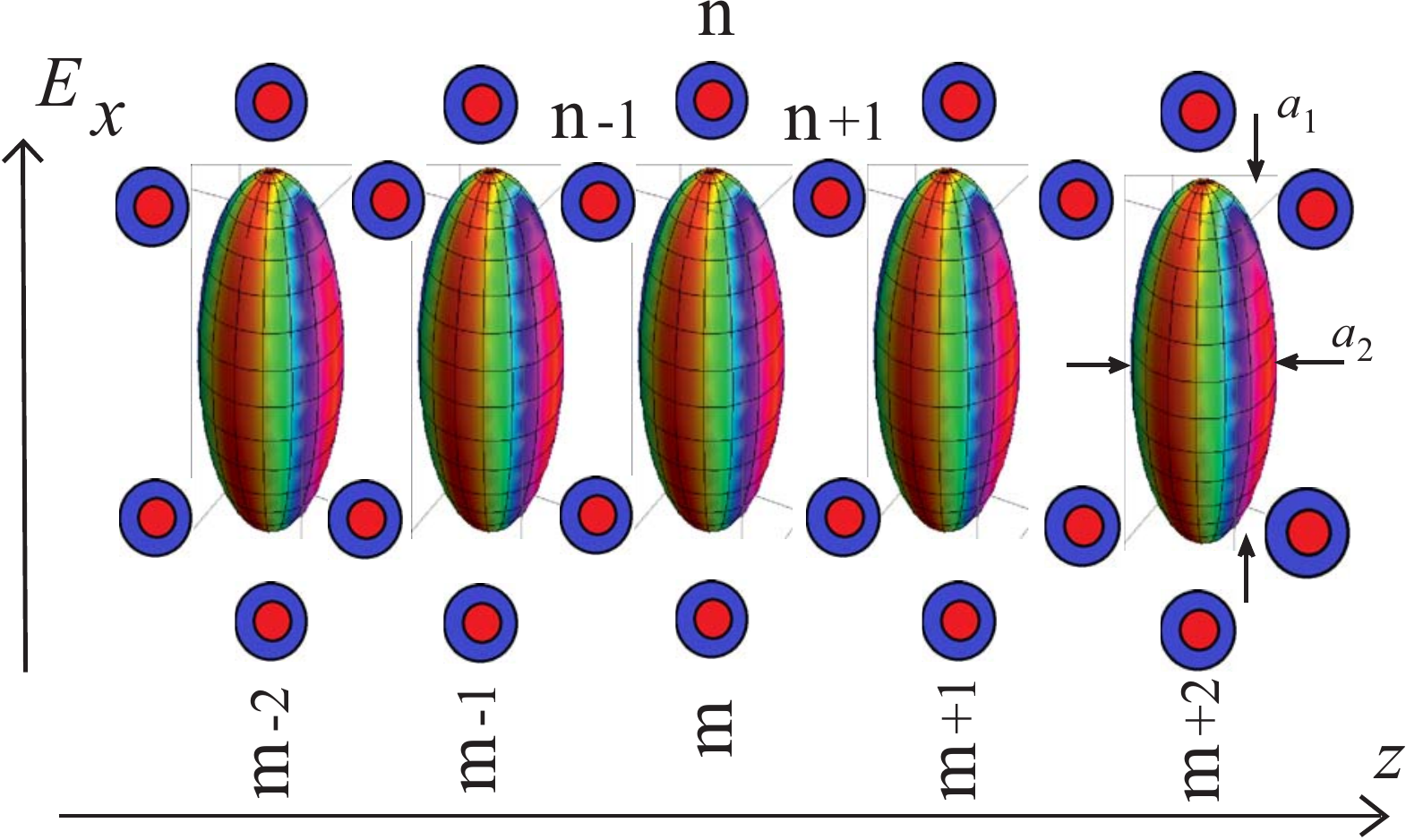} 
\caption{\label{fig2} Schematic representation of the 1D chain of the \textsl{NP}-\textsl{QD} used to describe the \textsl{NP}-\textsl{QD} chain in Fig.~\ref{fig1}. The yellow bricks represent \textsl{NP} (with the main axis $a_{1}$ and $a_{2}$), while active molecules (\textsl{QD}) are shown by small red circles.}
\end{figure}

We investigate the 1D array of plasmonic nano particles (\textsl{NP}) optically coupled with each other and with the quantum systems (see Fig.~\ref{fig1}). Hereafter, the \textsl{QD} are assumed as quantum systems. The \textsl{NP}s are assumed to be small enough to be adequately descried by a linear harmonic oscillator equation for the dipole moment amplitude $\tilde{p}_{m} \left(t\right)$: 
\begin{equation} \label{EQ__1_} 
\frac{d^{2} }{dt^{2} } \tilde{p}_{m} (t)+2\gamma \frac{d}{dt} \tilde{p}_{m} (t)+\omega _{\textsl{NP},m}^{2} \tilde{p}_{m} (t)=2\chi \tilde{E}_{m} (t).         
\end{equation} 
Here $\omega _{\textsl{NP},m}$ is the eigen \textsl{NP} frequency, $\gamma$ is the damping coefficient, $\chi $ is the susceptibility, $\tilde{E}_{m} $ is the total electric field acting on this \textsl{NP} which is basically sum of the external field and the field from other oscillators. We assume here layout shown in Fig.~\ref{fig2}, namely all \textsl{NP}s are aligned along the $x$ axis, while the 1D array is placed along the $z$ axis. In general, the \textsl{NP}s are anisotropic ones (e.g. have ellipsoid shape) and the values $\gamma$, 
$\omega _{\textsl{NP},m}$ and $\chi $ become tensors. In the presented here picture only $\tilde{p}_{x,m} \left(t\right)$ is of interest and equation \eqref{EQ__1_} has to be considered as an equation for the $x$ dipole vector component. 

Slowly Varying Approximation (SVA) for the $x$ components of the \textit{m-th} dipole moment in the chain of the nanoresonators (nanoparticles - \textsl{NP}) and of the field is:
\begin{equation} \label{EQ__2_} 
\frac{dp_{m} }{dt} +\left(\gamma _{\textsl{NP},m} -i\delta \omega _{\textsl{NP},m} \right)p_{m} =i\chi _{\textsl{NP},m} A_{m} ,                          
\end{equation} 
\noindent here $\tilde{E}\left(t\right)=\text{Re}[A\left(t\right)e^{-i\omega t} ]e_{x} $,\\
$\tilde{p}_{m} \left(t\right)=\text{Re}[\varepsilon _{h} p_{m} \left(t\right)e^{-i\omega t} ]e_{x} $, $\delta \omega _{\textsl{NP},m} =\omega -\omega _{\textsl{NP},m} $, $\chi _{\textsl{NP},m} =\chi /\left(\varepsilon _{h} \omega _{\textsl{NP},m} \right)$, and $\gamma _{\textsl{NP},m} =\gamma /\varepsilon _{h} \omega _{\textsl{NP},m} $. The use of the harmonic oscillator equations for the description of the \textsl{NP} dynamics turned out to be extremely convenient and combine math simplicity with physical adequateness, making the respective models analytically (or semi analytically) treatable. From the other side the analytical results under certain restrictions correspond with the rigorous numerical and experimental data not only qualitatively, but quantitatively as well \cite{Petschulat08,Petschulat10,Chipouline11}.  

The internal quantum \textsl{QD} dynamics is described by the well-known Maxwell-Bloch pair with the artificial pump \cite{Chipouline12}. For the inversion $N_{nm} \left(t\right)$ (\textit{n-th} \textsl{QD} on \textit{m-th} \textsl{NP}, see Fig.~\ref{fig2}), complex \textsl{QD} dipole moment amplitude $R_{nm} \left(t\right)=\text{Re}[R_{nm} \left(t\right)e^{-i\omega t} ]$ ($p_{\textsl{QD},nm} =\mu _{\textsl{QD},n} R_{nm} \left(t\right)$ is the dipole momentum of \textsl{QD}):
\begin{equation} \label{EQ__3a_} 
\begin{array}{l} 
{\displaystyle\frac{dN_{nm} }{dt} +\gamma _{1,nm} \left(N_{nm} -N_{0nm} \right)=\frac{\mu _{\textsl{QD},nm} \text{Im}\left[ A_{nm}^{*} R_{nm} \right]
}{\hbar }\text{,} } \\[0.5\normalbaselineskip]
 {\displaystyle\frac{dR_{nm} }{dt} +\left(\gamma _{2,nm} -i\delta \omega _{\textsl{QD},nm} \right)R_{nm} =\frac{\mu _{\textsl{QD},nm} A_{nm} N_{nm}}{i\hbar }.} 
 \end{array}
\end{equation} 
\noindent Here $\mu _{\textsl{QD}} =\bar{\mu }_{\textsl{QD}} /\left(\varepsilon _{h} \varepsilon '_{\textsl{QD}} \right)$ is the reduced \textsl{QD} dipole moment, $\bar{\mu }_{\textsl{QD}} $ is the dipole moment of the transition, $\varepsilon _{h} $ is the host dielectric constant, $\varepsilon _{\textsl{QD}} $ is the dielectric constant of the \textsl{QD}, and $\varepsilon '_{\textsl{QD}} =\left(\varepsilon _{\textsl{QD}} +2\varepsilon _{h} \right)/3\varepsilon _{h} $ is the screening factor \cite{Malyshev11}; $\gamma _{1,nm} =\frac{2\widetilde{\gamma }_{1} }{1-N_{0,nm} } $, $N_{0,nm} =\frac{\left(W\, -\, \widetilde{\gamma }_{1} \right)}{\left(W\, +\, \widetilde{\gamma }_{1} \right)} $ and $\gamma _{2,nm} $ are the energy and phase relaxation rates respectively, $\delta \omega _{\textsl{QD},nm} =\omega -\omega _{\textsl{QD},nm} $ is the frequency detuning of the resonance of (\textit{n, m-th}) \textsl{QD} from the incident field carrier frequency. It is worth noting here, that in the case of the absence of the external field the frequency $\omega $ is the spaser generation frequency which is supposed to be found from the stationary solution, provided the pump is high enough to support the spaser generation. 

In order to develop the model further and include interaction between the \textsl{NP}s and \textsl{QD}s, the realistic picture shown in Fig.~\ref{fig2} has to be simplified. We substitute the \textsl{NP}s by the point dipoles, and all \textsl{QD}s placed on a \textsl{NP} we substitute by a single \textsl{QD} placed at the distance $r_{\textsl{QD}} $from the \textsl{NP}; in this case the second subscript of the \textsl{QD} polarization can be left $R_{nm}^{} \equiv R_{m}^{} $, $N_{nm} \equiv N_{m} $. Actually, the dynamics of this ``effective \textsl{QD}'' is described by the same density matrix/Maxwell-Bloch equations; from the other side, for the interaction between the ``effective \textsl{QD}'' and the \textsl{NP}s we have to introduce an effective number of the \textsl{QD}s placed on the \textit{m-th} \textsl{NP} $n_{\textsl{QD},m} $, and effective \textsl{QD} dipole momentum, $\mu _{\textsl{QD},m} =n_{\textsl{QD},m} \mu _{\textsl{QD}} $ (and the same effective dipole momentum in host medium, $\tilde{\mu}_{\textsl{QD},m} =\varepsilon _{h}\mu _{\textsl{QD},m} $, $\tilde{\mu}_{\textsl{QD}} =\varepsilon _{h}\mu _{\textsl{QD}} $). System of equation (\eqref{EQ__3a_}) is finally reduced to:
\begin{equation} \label{EQ__3b_} 
\begin{array}{l} 
{\displaystyle\frac{dN_{m} }{dt} +\gamma _{1,m} \left(N_{m} -N_{0m} \right)=\frac{i\tilde{\mu} _{\textsl{QD},m} \left(A_{m} R_{m}^{*} \, -\, A_{m}^{*} R_{m}^{} \right)}{2\hbar },} 
\\[0.5\normalbaselineskip] {\displaystyle\frac{dR_{m} }{dt} +\left(\gamma _{2,m} -i\delta \omega _{\textsl{QD},m} \right)R_{m} =-\frac{i\tilde{\mu} _{\textsl{QD},m} A_{m} N_{m}}{\hbar }.} \end{array}                                        
\end{equation} 
To this extend, the approximation is equivalent to the one accepted in \cite{Andrianov12}. It is worth noting again, that one of the main differences with \cite{Andrianov12} is in the model of interaction between the \textsl{NP}s in the chain. Namely, authors of \cite{Andrianov12} took into account only interaction with the neighboring \textsl{NP}s, which excludes appearance of the narrowband resonances. The narrowband resonances is the consequence of the interaction of each \textsl{NP} with all other \textsl{NP}s in the chain \cite{Markel05}; investigation of these resonances is the main goal of the presented here consideration.        

The interaction between the \textsl{NP}s is described in the frame of the dipole approximation for each \textsl{NP} and \textsl{QD}. Namely, the generated by each dipole (\textsl{NP} plus \textsl{QD}s) \textit{m} field at the point of the dipole (\textsl{NP} or \textsl{QD}) \textit{m'} is given by:
\begin{equation} \label{EQ__4_} 
\begin{array}{l} 
{\displaystyle A_{mm'} \left(\omega \right)= C\left(r_{mm'}\right)
 \left(p_{m'} +p_{\textsl{QD},m'} \right) e^{ik_{0} r_{mm'}},} \\[0.5\normalbaselineskip]
{\displaystyle C\left(r_{mm'}\right)=
\frac{k_{0}^{2} }{r_{mm'} } +\frac{ik_{0} }{r_{mm'}^{2} } -\frac{1}{r_{mm'}^{3} }. }
\end{array}
\end{equation}
Here $k_{0} $is the wave vector of the field in space between the \textsl{NP}s, $r_{mm'} $ is the distance between the \textsl{NP}s, and $p_{m'} $ and $p_{\textsl{QD},m'} =\mu _{\textsl{QD},m'} R_{m'} $ are the complex amplitudes of the dipole moment of \textit{m'-th} \textsl{NP} and \textsl{QD}s placed on this \textsl{NP} respectively. Each \textsl{NP}/\textsl{QD} is driven by the field resulted by the sum of an external field $A_{ext,m} (t)$and all fields from the other \textsl{NP}s/\textsl{QD}s:
\begin{equation} \label{EQ__5_} 
A_{m} (\omega )=A_{ext,m} (\omega )+\sum _{m'}A_{mm'} (\omega ).
\end{equation} 
It is convenient to introduce more concrete equations for the sums involved into the consideration. Taking into account, that the self-action (action of the \textsl{NP}/\textsl{QD} on itself) has to be excluded, we introduce ``sums of interaction'', namely: 
\begin{equation} \label{EQ__6_} \begin{array}{lcr} 
\sum _{m'}A_{mm'} (\omega )& = & A_{m,\textsl{NP}} \left(\omega \right)+A_{m,\textsl{NP},\textsl{QD}} \left(\omega \right)+ \\ &  & +
A_{m,\textsl{QD},\textsl{NP}} \left(\omega \right)+A_{m,\textsl{QD}} \left(\omega \right) 
 \end{array}
\end{equation}  
\begin{equation} \label{EQ__7_} 
\begin{array}{l} {A_{m,\textsl{NP}} \left(\omega \right)=\sum _{m\ne m'}
C\left(r_{mm'}\right)p_{m'} \left(\omega \right)e^{ik_{0} r_{mm'} },  } \\ {A_{m,\textsl{NP},\textsl{QD}} \left(\omega \right)=\sum _{m'}
C\left(r_{mm'}\right)p_{\textsl{QD},m'} \left(\omega \right)e^{ik_{0} r_{mm'} },  } \\ {A_{m,\textsl{QD},\textsl{NP}} \left(\omega \right)=\sum _{m'}
C\left(r_{mm'}\right)p_{m'} \left(\omega \right)e^{ik_{0} r_{mm'} },  } \\ 
{A_{m,\textsl{QD}} \left(\omega \right)=\sum _{m\ne m'}
C\left(r_{mm'}\right)p_{\textsl{QD},m'} \left(\omega \right)e^{ik_{0} r_{mm'} },  } \\  {A_{m} \left(\omega \right)=A_{ext,m} \left(\omega \right)+A_{m,\textsl{NP}} \left(\omega \right)+}\\ \quad 
{+A_{m,\textsl{NP},\textsl{QD}} \left(\omega \right)+A_{m,\textsl{QD},\textsl{NP}} \left(\omega \right)+A_{m,\textsl{QD}} \left(\omega \right).} \end{array} 
\end{equation} 
The final field $A_{m} (\omega )$ acting on \textit{m-th} \textsl{NP} and \textsl{QD} (see (5)) is given by the sum of all the partial sums -- see last equation in \eqref{EQ__7_}. 

The physical means of the partial sums are clear: $A_{m,\textsl{NP}} \left(\omega \right)$is the field acting on the \textit{m-th} \textsl{NP} from the other \textsl{NP}s, $A_{m,\textsl{NP},\textsl{QD}} \left(\omega \right)$ is the field acting on the \textit{m-th} \textsl{NP} from all \textsl{QD}s, $A_{m,\textsl{QD},\textsl{NP}} \left(\omega \right)$ is the field acting on the \textit{m-th} \textsl{QD} from all \textsl{NP}s, and $A_{m,\textsl{QD}} \left(\omega \right)$ is the field acting on the \textit{m-th} \textsl{QD} from all other \textsl{QD}s. 

Equations (2-7) form the closed system for the coupled dynamics of the coupled \textsl{NP}-\textsl{QD} array. These equations will be used to investigate the stationary states of the considered system depending on the parameters, including the level of pump $N_{0,m} $. At the absence of pump $N_{0,m} =-1$, at $-1<N_{0,m} <0$ the \textsl{QD}s still absorb, and at  $N_{0,m} >0$ starts to amplify the light. 

Coupled system of the \textsl{NP}-\textsl{QD} is known as spaser (one of the possible realization). Dynamics of the spaser is described by the same system of equations (2-7) without an external field
 $A_{ext} (t)=0$. Respective solutions depend on the pump level: it is zero for the pump values below and nonzero for the pumps above some threshold value $N_{th} $ which is supposed to be found from the solution of system (2-7). In the case of the nonzero external field $A_{ext} (t)\ne 0$ the nonzero solutions exist in both below and above threshold regions; moreover the system possess several solutions manifesting the multistable operation \cite{Andrianov11}.  

Here we are interested primarily in solutions for the pump levels below lasing threshold. These values are realized in most experiments, and the resulted response of such system (amplitude and phase) is actually most easily can be compared with the experimental data. Note that the spontaneous emission is included in the model by the energy relaxation rate $\gamma _{1} $, which is subject to modification by Purcell effect. The relaxation dynamics in this coupled system can be also investigated in the frame of the presented model which is left for further publications.  

\section{\label{sec:HomArr}Homogeneous 1D array}

We assume for simplicity that all the \textsl{NP}s and \textsl{QD} have the same properties (in (2-7) the respective coefficients lose their subscript \textit{m}). System of equations (2-7) in this case becomes:
\begin{equation} \label{EQ__8_} 
\begin{array}{l} { \displaystyle
\frac{dN_{m} }{dt} +\gamma _{1} \left(N_{m} -N_{0} \right)=\frac{i\tilde{\mu}_{\textsl{QD}}}{2\hbar }
\left( A_{\textsl{QD}}^{(m)}R_{m}^{*}-A_{\textsl{QD}}^{(m)*} R_{m} \right),} 
\\[0.5\normalbaselineskip] { \displaystyle
\frac{dR_{m} }{dt} +\left(\gamma _{2} -i\delta \omega _{\textsl{QD}} \right)R_{m} 
 = -\frac{i\tilde{\mu}_{\textsl{QD}}}{\hbar }A_{\textsl{QD}}^{(m)} N_{m},
} \\[0.5\normalbaselineskip] { \displaystyle
\frac{dp_{m} }{dt} +\left(\gamma _{\textsl{NP}} -i\delta \omega _{\textsl{NP}} \right)p_{m} 
 = i\chi _{\textsl{NP}} A_{\textsl{NP}}^{(m)},
} \end{array} 
\end{equation}
where \begin{equation*} 
 \begin{array}{l} {\delta \omega _{\textsl{QD}} =\omega -\omega _{\textsl{QD}}, \;
 A_{\textsl{QD}}^{(m)}=A_{ext,m} +A_{m,\textsl{QD},\textsl{NP}} +A_{m,\textsl{QD}},
 } \\  {\delta \omega _{\textsl{NP}} =\omega -\omega _{\textsl{NP}}, \;
 A_{\textsl{NP}}^{(m)}=A_{ext,m} +A_{m,\textsl{NP}}  + A_{m,\textsl{NP},\textsl{QD} }.
  } \end{array} \end{equation*}
In this paper the consideration is restricted by the stationary operation mode with the respective stationary solution given by zeroing all derivatives in \eqref{EQ__8_}:
\begin{equation} \label{EQ__9_} 
\begin{array}{l} { \displaystyle
N_{m} =N_{0} +\frac{i\tilde{\mu}_{\textsl{QD}} }{2\hbar \gamma _{1} } \left( A_{\textsl{QD}}^{(m)}R_{m}^{*}-A_{\textsl{QD}}^{(m)*} R_{m} \right),} \\[0.5\normalbaselineskip]  { \displaystyle
R_{m} =-\frac{i\tilde{\mu}_{\textsl{QD}} }{\hbar \left(\gamma _{2} -i\delta \omega _{\textsl{QD}} \right)} A_{\textsl{QD}}^{(m)}N_{m},} \\[0.5\normalbaselineskip]  { \displaystyle
p_{m} =\frac{i\chi _{\textsl{NP}} }{\left(\gamma _{\textsl{NP}} -i\delta \omega _{\textsl{NP}} \right)} A_{\textsl{NP}}^{(m)}.
} \end{array} 
\end{equation} 
System \eqref{EQ__9_} can be further simplified. It is assumed, that the \textsl{QD}s interact mainly with their ``own'' resonators, which reduces the sums $A_{m,\textsl{NP},\textsl{QD}} \left(\omega \right)$ and $A_{m,\textsl{QD},\textsl{NP}} \left(\omega \right)$ in \eqref{EQ__7_} to only one term and sets $A_{m,\textsl{QD}} \left(\omega \right)=0$:  
\begin{equation} \label{EQ__10a_} 
\begin{array}{l} {A_{m,\textsl{NP},\textsl{QD}} \left(\omega \right)=\sum _{m'}
C\left(r_{mm'}\right)p_{\textsl{QD},m'} \left(\omega \right)e^{ik_{0} r_{mm'} }  =
} \\[0.5\normalbaselineskip] {\, \, = C\left(r_{\textsl{QD}}\right)
p_{\textsl{QD},m} \left(\omega \right)e^{ik_{0} r_{mm} } \approx 
-\mu _{\textsl{QD}} R_{m} \left(\omega \right)/ r_{\textsl{QD}}^{3},
 } \end{array}
\end{equation} 
\begin{equation} \label{EQ__10b_} 
\begin{array}{l} {A_{m,\textsl{QD},\textsl{NP}} \left(\omega \right)=\sum _{m'}
C\left(r_{mm'}\right)p_{m'} \left(\omega \right)e^{ik_{0} r_{mm'} }  =
} \\[0.5\normalbaselineskip] { \, \, = C\left(r_{\textsl{QD}}\right)
p_{m} \left(\omega \right)e^{ik_{0} r_{mm'} } \approx p_{m} / r_{\textsl{QD}}^{3}.
 } \end{array}
\end{equation}
Assuming that the external field is a plane wave incoming at the angle $\theta $ to the perpendicular to the plane with the \textsl{NP} chain, see Fig.~\ref{fig1}, which makes for the $A_{ext,m} \left(\omega \right)$: 
\begin{equation} \label{EQ__11_} 
A_{ext,m} \left(\omega \right)=A_{ext} \left(\omega \right)\exp \left(ik_{0} md\sin \left(\theta \right)\right). 
\end{equation} 
Using Bloch theorem, assume also the periodic dependence on $z$ for all variables in \eqref{EQ__9_}, namely:
\begin{equation} \label{EQ__12_} 
\begin{array}{l} {N_{m} \left(\omega \right)=N\left(\omega ,k\right)\exp \left(ikmd\right),} \\
{R_{m} \left(\omega \right)=R\left(\omega ,k\right)\exp \left(ikmd\right),} \\ {p_{m} \left(\omega \right)=p\left(\omega ,k\right)\exp \left(ikmd\right).} \end{array} 
\end{equation} 
$k$ is the $z$ component of the Bloch vector which allows us in turn to simplify the sums $A_{m,\textsl{NP}}$ and $A_{m,\textsl{QD}}$ in \eqref{EQ__7_}: \vspace{-6pt}
\begin{equation} \label{EQ__13_} 
\begin{array}{l} {A_{m,\textsl{NP}} \left(\omega ,k\right)=S_{1} \left(\omega ,k\right)p\left(\omega ,k\right),} \\[0.5\normalbaselineskip] {S_{1} \left(\omega ,k\right)=2\, \sum _{m=1}^{\infty }
C\left(dm\right)e^{ik_{0} dm} \cos \left(kdm\right).} \end{array} 
\end{equation} \vspace{6pt}

Substituting (10-13) into \eqref{EQ__9_}, the system of equations for stable state operation mode becomes:
\begin{widetext}
\begin{equation} \label{EQ__14_}  \left\{
\begin{array}{l} { \displaystyle
  N\left(\omega ,k\right)\exp \left(ikmd\right) = 
  N_{0} +\frac{i\tilde{\mu}_{\textsl{QD}} }{2\hbar \gamma _{1} } \left(
   \left(A_{ext}\left(\omega,k \right)-\frac{p\left(\omega ,k\right)}{r_{\textsl{QD}}^{3}} \right)
  R^{*}\left(\omega ,k\right) - \left(A_{ext}\left(\omega,k \right)-
   \frac{p\left(\omega ,k\right)}{r_{\textsl{QD}}^{3}} \right)^{*} R \left(\omega ,k\right) \right),
 } \\[\normalbaselineskip] { \displaystyle
  R\left(\omega ,k\right) =
  -\frac{i\tilde{\mu}_{\textsl{QD}} }{\hbar \left(\gamma _{2} -i\delta \omega _{\textsl{QD}} \right)} 
 \left(A_{ext}\left(\omega,k \right)-\frac{p\left(\omega ,k\right)}{r_{\textsl{QD}}^{3}} \right)
 N\left(\omega ,k\right)\exp \left(ikmd\right), } \\[\normalbaselineskip] { \displaystyle 
 p\left(\omega ,k\right)= 
  \frac{i\chi _{\textsl{NP}} }{\left(\gamma _{\textsl{NP}} -i\delta \omega _{\textsl{NP}} \right)} 
\left(A_{ext}\left(\omega,k \right)+p\left(\omega ,k\right)S_{1} \left(\omega ,k\right)-
 \frac{\mu _{\textsl{QD}} R\left(\omega ,k\right)}{r_{\textsl{QD}}^{3}} \right).}
 \end{array} \right.
\end{equation}
\end{widetext}
where $A_{ext}\left(\omega,k \right) = A_{ext} \left(\omega \right)\exp \left(imd\left(k_{0} \sin \left(\theta \right)-k\right)\right) $ and $S_{1} \left(\omega ,k\right)$ is dipole sum in (\eqref{EQ__13_}).

At this stage system possesses pretty clear physical interpretation. Let us suppose first absence of the external field $A_{ext} =0$. In this case the considered system is a combination of coupled nanolasers/spasers. It is clear, that there are two possibilities for the steady states: in case of low pump ($N\left(\omega ,k\right)<N_{th} $, where $N_{th}$ is the threshold inversion i.e. minimum inversion necessary for the generation) the nanolaser/spaser cannot reach nonzero steady state at any frequency $\omega $  ($p\left(\omega ,k\right)=0$ and $R\left(\omega ,k\right)=0$). In the case of high enough pumps ($N\left(\omega ,k\right)>N_{th} $), the nanolaser/spaser after some transition period reaches the steady state which is characterized by nonzero values of $p\left(\omega _{G} ,k_{G} \right)\ne 0$ and $R\left(\omega _{G} ,k_{G} \right)\ne 0$; moreover, the generation frequency $\omega _{G} $ and $k_{G} $ is the eigen frequency and mode eigen wave vector of the steady state, determined by the solution of \eqref{EQ__15_}:
\begin{widetext}
\begin{equation} \label{EQ__15_} 
\left\{\begin{array}{l} { \displaystyle
 N\left(\omega _{G} ,k_{G} \right)\exp \left(ik_{G} md\right)=N_{0} +\frac{i\tilde{\mu }_{\textsl{QD}} }{2\hbar \gamma _{1} r_{\textsl{QD}}^{3} } \left(p\left(\omega _{G} ,k_{G} \right)^{*} R\left(\omega _{G} ,k_{G} \right)-p\left(\omega _{G} ,k_{G} \right)R^{*} \left(\omega _{G} ,k_{G} \right)\, \right),}  \\[0.5\normalbaselineskip] { \displaystyle
R\left(\omega _{G} ,k_{G} \right)=\frac{i\tilde{\mu }_{\textsl{QD}} }{\hbar r_{\textsl{QD}}^{3} \left(\gamma _{2} -i\left(\omega _{G} -\omega _{\textsl{QD}} \right)\right)} p\left(\omega _{G} ,k_{G} \right)N\left(\omega _{G} ,k_{G} \right)\exp \left(ik_{G} md\right),
 } \\[0.5\normalbaselineskip]  { \displaystyle
 p\left(\omega _{G} ,k_{G} \right)=\frac{i\chi _{\textsl{NP}} }{\left(\gamma _{\textsl{NP}} -i\left(\omega _{G} -\omega _{\textsl{NP}} \right)\right)} \left(p\left(\omega _{G} ,k_{G} \right)S_{1} \left(\omega _{G} ,k_{G} \right)-\frac{\mu _{\textsl{QD}} R\left(\omega _{G} ,k_{G} \right)}{r_{\textsl{QD}}^{3} } \right),} \\[\normalbaselineskip] {
 S_{1} \left(\omega _{G} ,k_{G} \right)=2\, \sum _{m=1}^{\infty }\left(\frac{k_{0}^{2} }{dm} +\frac{ik_{0} }{d^{2} m^{2} } -\frac{1}{d^{3} m^{3} } \right)e^{ik_{0} dm} \cos \left(k_{G} dm\right) .} \end{array}\right.  
\end{equation} 
Expressing $p\left(\omega _{G} ,k_{G} \right)$ from the last and substituting into the first equation of \eqref{EQ__15_}, we get:
\begin{equation} \label{EQ__16_} 
\left\{\begin{array}{l} { \displaystyle
N\left(\omega _{G} ,k\right)\exp \left(ik_{G} md\right)=} \\[0.5\normalbaselineskip] { \displaystyle
 \quad =N_{0} -\frac{\tilde{\mu }_{\textsl{QD}}^{2} \chi _{\textsl{NP}} }{\hbar \gamma _{1} r_{\textsl{QD}}^{6} } \frac{\left(\gamma _{\textsl{NP}} +\chi _{NP} \text{Im}\left[S_{1} \left(\omega _{G} ,k_{G} \right)\right]\right)}{\left(\gamma _{NP} +\chi _{NP} \text{Im}\left[S_{1} \left(\omega _{G} ,k_{G} \right)\right]\right)^{2} +\left(\omega _{G} -\omega _{\textsl{NP}} +\chi _{\textsl{NP}} \text{Re}\left[S_{1} \left(\omega _{G} ,k_{G} \right)\right]\right)^{2} } \left|R\left(\omega _{G} ,k_{G} \right)\right|^{2}, } \\[\normalbaselineskip]  { \displaystyle
p\left(\omega _{G} ,k_{G} \right)=\frac{-i\chi _{\textsl{NP}} \mu _{\textsl{QD}}}{r_{\textsl{QD}}^{3} \left(\gamma _{\textsl{NP}} +\chi _{\textsl{NP}} \text{Im}\left[S_{1} \left(\omega _{G} ,k_{G} \right)\right]-i\left(\omega _{G} -\omega _{\textsl{NP}} +\chi _{\textsl{NP}} \text{Re}\left[S_{1} \left(\omega _{G} ,k_{G} \right)\right]\right)\right)} R\left(\omega _{G} ,k_{G} \right). } \end{array}\right.  
\end{equation} 
Substituting \eqref{EQ__16_} into the second equation of \eqref{EQ__15_}, we obtain two transcendental equations for eigen frequency $\omega _{G} $,  mode eigen wave vector $k_{G} $, and squares of the amplitudes of the stationary oscillations $\left|R\left(\omega _{G} ,k_{G} \right)\right|^{2} $and $\left|p\left(\omega _{G} ,k_{G} \right)\right|^{2} $: 
\begin{equation} \label{EQ__17_} 
\left\{\begin{array}{l} { \displaystyle
\gamma _{2} =\frac{\tilde{\mu }_{\textsl{QD}}^{2} \chi _{\textsl{NP}} }{\hbar \gamma _{1} r_{\textsl{QD}}^{6} } \frac{\gamma _{\textsl{NP}} +\chi _{\textsl{NP}} 
\text{Im}\left[S_{1} \left(\omega _{G} ,k_{G} \right)\right]}{\left(\gamma _{\textsl{NP}} +\chi _{\textsl{NP}} \text{Im}\left[S_{1} \left(\omega _{G} ,k_{G} \right)\right]\right)^{2} +\left(\omega _{G} -\omega _{\textsl{NP}} +\chi _{\textsl{NP}} \text{Re}\left[S_{1} \left(\omega _{G} ,k_{G} \right)\right]\right)^{2} } \Delta N,
} \\[\normalbaselineskip] { \displaystyle
\omega _{G} -\omega _{\textsl{QD}} =\frac{\tilde{\mu }_{\textsl{QD}}^{2} \chi _{\textsl{NP}} }{\hbar \gamma _{1} r_{\textsl{QD}}^{6} } \frac{\omega _{G} -\omega _{\textsl{NP}} +\chi _{\textsl{NP}} \text{Re}\left[S_{1} \left(\omega _{G} ,k_{G} \right)\right]}{\left(\gamma _{\textsl{NP}} +\chi _{\textsl{NP}} \text{Im}\left[S_{1} \left(\omega _{G} ,k_{G} \right)\right]\right)^{2} +\left(\omega _{G} -\omega _{\textsl{NP}} +\chi _{\textsl{NP}} \text{Re}\left[S_{1} \left(\omega _{G} ,k_{G} \right)\right]\right)^{2} } \Delta N,
} \\[\normalbaselineskip] { \displaystyle
\Delta N=N_{0} -\frac{\tilde{\mu }_{\textsl{QD}}^{2} \chi _{\textsl{NP}} }{\hbar \gamma _{1} r_{\textsl{QD}}^{6} } \frac{\left(\gamma _{\textsl{NP}} +\chi _{\textsl{NP}} \text{Im}\left[S_{1} \left(\omega _{G} ,k_{G} \right)\right]\right)\left|R\left(\omega _{G} ,k_{G} \right)\right|^{2} }{\left(\gamma _{\textsl{NP}} +\chi _{\textsl{NP}} \text{Im}\left[S_{1} \left(\omega _{G} ,k\right)\right]\right)^{2} +\left(\omega _{G} -\omega _{\textsl{NP}} +\chi _{\textsl{NP}} \text{Re}\left[S_{1} \left(\omega _{G} ,k_{G} \right)\right]\right)^{2}. } } \end{array}\right.  
\end{equation} 
\end{widetext}
In case of absence of the interaction between the \textsl{NP}s ($S_{1} \left(\omega _{G} ,k_{G} \right)=0$) \eqref{EQ__17_} gives the standard equations for eigen frequency and amplitude of the spaser generation; in case of the interaction only with the neighborings \eqref{EQ__17_} describes the stable state in the interacting spaser chain considered in \cite{Andrianov11}. It is worth noting that the used here (and in \cite{Andrianov11}) approach does not allow us to investigate the laser bandwidth, which requires stochastic methods \cite{Chirkin11}.  

Full investigation of \eqref{EQ__17_} is out of the scopes of this paper. The question about which spatial modes $k_{G} $ can reach generation depends on the level of pump $N_{0} $; usually the mode with the minimum threshold appears first. For example, dark modes (with the effective zero dipole moment) have less radiative losses (and consequently lower total losses) and should appear first. The developed here model does not take into account the specific for different modes radiative losses (it has to be included as an extra losses which depend on $k$) and therefore is not appropriate for the full analysis. Below, only the first spatial mode ($k_{G} =0$) of the chain will be considered in order to find the generation threshold and distinguish between the operation modes below and above the generation threshold. 

The physical picture of the stationary dynamics is different in case of a nonzero external field $A_{ext} \ne 0$ . In this case $k$ is basically fixed $k=k_{0} \sin \left(\theta \right)$ and system \eqref{EQ__14_} for stationary solution becomes:
\begin{widetext}
\begin{equation} \label{EQ__18_} 
\left\{\begin{array}{l} { \displaystyle
N\left(\omega ,k\right)e^{ikmd} =N_{0} +\frac{i\tilde{\mu }_{\textsl{QD}} }{2\hbar \gamma _{1} } \left(\left(A_{ext} \left(\omega \right)-\frac{p\left(\omega ,k\right)}{r_{\textsl{QD}}^{3} } \right)R^{*} \left(\omega ,k\right)\, -\, \left(A_{ext} \left(\omega \right)-\frac{p\left(\omega ,k\right)}{r_{\textsl{QD}}^{3} } \right)^{*} R\left(\omega ,k\right)\right),
} \\[0.5\normalbaselineskip] { \displaystyle
R\left(\omega ,k\right)=-\frac{i\tilde{\mu }_{\textsl{QD}} }{\hbar \left(\gamma _{2} -i\delta \omega _{\textsl{QD}} \right)} \left(A_{ext} \left(\omega \right)-\frac{p\left(\omega ,k\right)}{r_{\textsl{QD}}^{3} } \right)N\left(\omega ,k\right)\exp \left(ikmd\right),
} \\[0.5\normalbaselineskip] { \displaystyle
p\left(\omega ,k\right)=\frac{i\chi _{\textsl{NP}} }{\left(\gamma _{\textsl{NP}} -i\delta \omega _{\textsl{NP}} \right)} \left(A_{ext} \left(\omega \right)+p\left(\omega ,k\right)S_{1} \left(\omega ,k\right)-\frac{\mu _{\textsl{QD}} R\left(\omega ,k\right)}{r_{\textsl{QD}}^{3} } \right),
} \\[0.5\normalbaselineskip] { \displaystyle
S_{1} \left(\omega ,k\right)=2\, \sum _{m=1}^{\infty }\left(\frac{k_{0}^{2} }{dm} +\frac{ik_{0} }{d^{2} m^{2} } -\frac{1}{d^{3} m^{3} } \right)e^{ik_{0} dm} \cos \left(kdm\right), } \\ {k=k_{0} \sin \left(\theta \right),} \end{array}\right.  
\end{equation} 
\end{widetext}
Physically this case has to be subdivided by two sub problems, namely response of the system with the pump level below and above the generation threshold. It has to be emphasized that these two situations are principally different: in the first case the system is basically passive and it is possible to calculate its spectrum (own resonances and their bandwidths), while in the second case the system is governed by own self consistent dynamics, which is to some extend independent from the external field. The problem of the resonance bandwidths becomes much more sophisticated and requires stochastic methods. As it was mentioned in \cite{Andrianov11}, the problem is equivalent to the well-known one of the dynamics of macro laser under the action of a resonant external field \cite{Pikovsky01}. 

In this paper, only perpendicular incidence is supposed, namely $\sin \left(\theta \right)=0$ which implies$k=0$ or, in other words, all \textsl{NP}s in the chain oscillate in phase. It is worth noting that in this case the system response is sought at the frequency of the external field (there is no eigen frequency). In this case system \eqref{EQ__18_} is simplified:
\begin{equation} \label{EQ__19_} 
\left\{\begin{array}{l} { \displaystyle
N\left(\omega \right)=N_{0} +\frac{i\tilde{\mu }_{\textsl{QD}} }{2\hbar \gamma _{1} } \text{Im}\left[
\left(A_{ext} \left(\omega \right)-\frac{p\left(\omega \right)}{r_{\textsl{QD}}^{3} }
  \right)R^{*} \left(\omega \right) \right],
} \\[0.5\normalbaselineskip] { \displaystyle
R\left(\omega \right)=-\frac{i\tilde{\mu }_{\textsl{QD}} }{\hbar \left(\gamma _{2} -i\delta \omega _{\textsl{QD}} \right)} \left(A_{ext} \left(\omega \right)-\frac{p\left(\omega \right)}{r_{\textsl{QD}}^{3} } \right)N\left(\omega \right),
} \\[0.5\normalbaselineskip] { \displaystyle
p\left(\omega \right)=\alpha _{\textsl{NP}} \left(A_{ext} \left(\omega \right)-\frac{\mu _{\textsl{QD}} R\left(\omega \right)}{r_{\textsl{QD}}^{3} } \right), } \end{array}\right.  
\end{equation}
where definition of polarizability of metallic chain support dipole sum in (\eqref{EQ__13_}):
\begin{equation*}
\begin{array}{l} { \displaystyle 
\alpha _{\textsl{NP}} =\frac{i\chi _{\textsl{NP}} }{\left(\gamma _{\textsl{NP}} -i\delta \omega _{\textsl{NP}} \right)\left(1-\frac{i\chi _{\textsl{NP}} }{\left(\gamma _{\textsl{NP}} -i\delta \omega _{\textsl{NP}} \right)} S_{1} \left(\omega \right)\right)} ,
} \\[0.1\normalbaselineskip] {
S_{1} \left(\omega \right)=2\, \sum _{m=1}^{\infty }\left(\frac{k_{0}^{2} }{dm} +\frac{ik_{0} }{d^{2} m^{2} } -\frac{1}{d^{3} m^{3} } \right)e^{ik_{0} dm}.
} \end{array}
\end{equation*}
Behavior of the sums like $S_{1} \left(\omega \right)$in \eqref{EQ__19_} has been investigated in \cite{Markel05}. The logarithmic divergence of these sums at $k_{0} d=2\pi n$, (where $n$is an integer number) is the reason for the narrowband resonances appearing in the framework of the elaborated model. 

\section{\label{sec:Rslts}Results representation}

System \eqref{EQ__19_} has been investigated numerically for the case of the absence of an external field $A_{ext} \left(\omega \right)=0$(coupled spasers) and for the case of the external field driving the chain. Throughout the paper, the results will be presented for zero pump ($N_{0} =-1$) and for the pump values below ($N_{0} <\, N_{th} $) and above ($N_{0} >\, N_{th} $) the generation threshold. 

To be precise, the numerical values have been taken similar to \cite{Petschulat08} for a single plasmonic nanoresonator. It is assumed that the axis relation of the spheroidal \textsl{NP}s in the chain is $a_{1} /a_{2} =11,7$ (see Fig.~\ref{fig2}), averaged radius is $a_{\textsl{NP}} =\sqrt[{3}]{a_{1}^{2} a_{2} } =20\, {\rm nm}$, which gives \textsl{NP} eigen resonance at $\lambda _{\textsl{NP}} =801,1\, {\rm nm}$. The resonance bandwidth taking into account the respective radiation losses is 
\[\gamma _{\textsl{NP}} \left(\omega _{\textsl{NP}} \right)\cong \gamma _{\textsl{NP}}^{(0)} +\frac{2}{3} \left(ka_{\textsl{NP}} \right)^{3} \tilde{\chi }_{\textsl{NP}} ,\quad\tilde{\chi }_{\textsl{NP}} =\chi _{\textsl{NP}} /a_{\textsl{NP}}^{3} ,\] 
where we use numerical values to be in keeping with \cite{Petschulat08}, $\tilde{\chi }_{\textsl{NP}} =6.71\, \, {\rm fs}^{-1} $, $\gamma _{\textsl{NP}} =0.083\, \, {\rm fs}^{-1} $($\gamma _{\textsl{NP}}^{(0)} =0.0315\, \, {\rm fs}^{-1} $) for the wavelength of the resulted narrowband resonance $\lambda _{\textsl{NBR}} =840,5\, {\rm nm}$.

The distance between the \textsl{NP}-\textsl{QD} cells was chosen to be 583,7 nm. Only first diffractive resonance has been considered $k_{0} d=2\pi $, $k_{0} =\sqrt{\varepsilon _{h} } 2\pi /\lambda _{{\rm 0}} $, $\varepsilon _{h} =2.07$ ($n_{h} =1.44$), which results in the dipole resonances in the chain $\lambda _{\textsl{NBR}} =840,5\, {\rm nm}$. The numerical values have been chosen in order to match the resonance wavelength $\lambda _{\textsl{NBR}} =840,5\, {\rm nm}$ with the respective one from \cite{Petschulat08}.   

The \textsl{QD} dipole moment was assumed to be $\bar{\mu }_{\textsl{QD}} =25\, \, Debay$ and the number of the \textsl{QD}s per \textsl{NP} is $n_{\textsl{QD}} =200$.  The \textsl{QD}s are supposed to be a homogeneously broadened with the phase relaxation time $\gamma _{2}^{-1} =100\, \, {\rm fs}$ which is comparable with the plasmonic resonance bandwidth. In order to provide an effective interaction, the \textsl{QD}s have to have a transition at the wavelength of the maximum interaction between \textsl{NP} and \textsl{QD}, which appears (as it will be seen later) at $\lambda _{\textsl{QD}} =841.2\, {\rm nm}$. The distance between \textsl{QD} and nanoparticles, which determines the interaction strength is $r_{\textsl{QD}} =2a_{\textsl{NP}} =40\, {\rm nm}$. Here $\tilde{\gamma }_{1}^{-1} =0.8\, {\rm ns}$ is the spontaneous emission rate (both radiative and non-radiative).

\begin{figure}[b]
\includegraphics[width=0.4\textwidth]{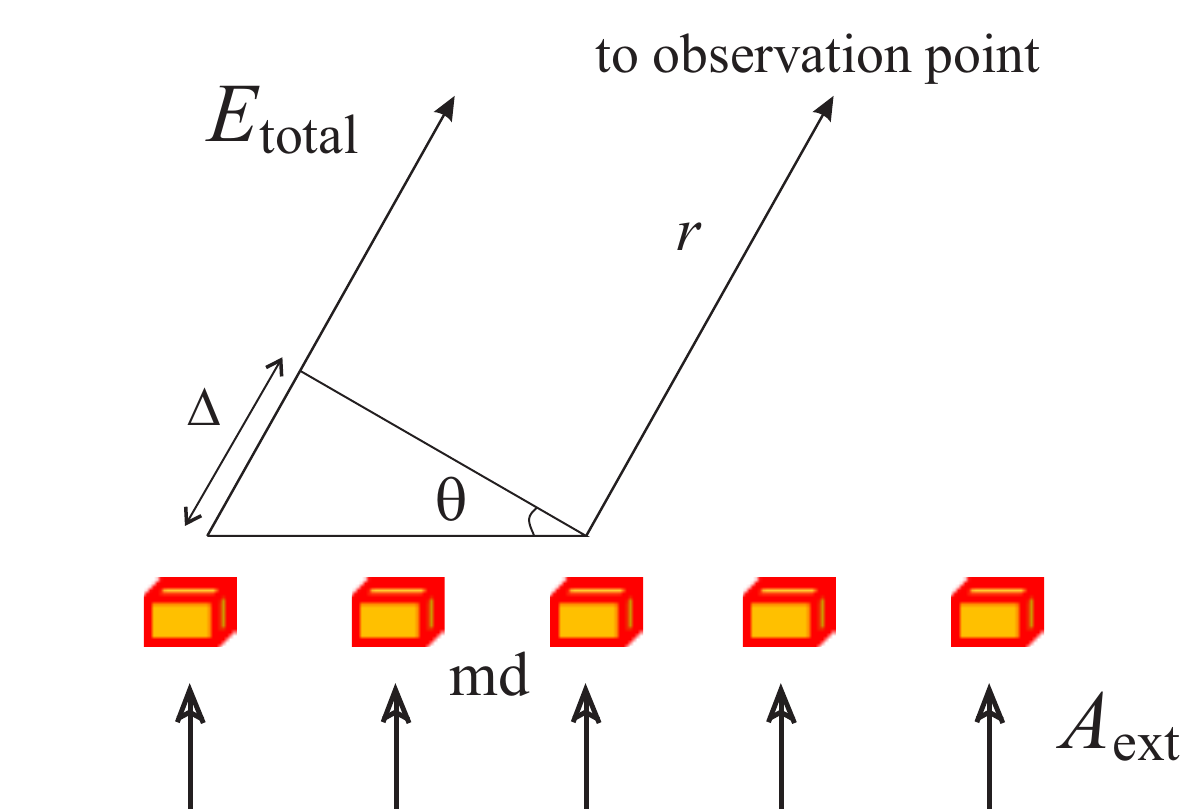} 
\caption{\label{fig3} Grating type contribution to the total field at the observation point. Gold bricks with red perimeters depict \textsl{NP}-\textsl{QD} cells, $\Delta =k_{0} md\, sin\left(\theta \right)$ is the pass difference contribution to the phase difference at the total field calculation. } \end{figure}

\noindent The system response has been investigated in the far field region. It is resulted by a coherent summation of the fields generated by each dipole $\mu _{\textsl{QD}} R\left(\omega \right)$ and $p\left(\omega \right)$ in both cases of presence or absence of the external field $A_{ext} \left(\omega \right)$. Frequency dependences of both functions $\mu _{\textsl{QD}} R\left(\omega \right)$ and $p\left(\omega \right)$ are given by \eqref{EQ__19_}, and the total dipole moment of the 
\textsl{NP}-\textsl{QD} is:
\begin{equation} \label{EQ__20_} 
p_{total} =p\left(\omega \right)+\mu _{\textsl{QD}} R\left(\omega \right) 
\end{equation} 
\noindent In order to calculate the total field from the chain (see Fig.~\ref{fig3}) the contributions from each \textsl{NP}-\textsl{QD} cell have to be summarized, see Fig.~\ref{fig3}. Each \textsl{NP}-\textsl{QD} contributes to the total field with own phase, namely: 
\begin{equation} \label{EQ__21_} 
E_{m,\textsl{NP}-\textsl{QD}} \left(\omega \right)=\frac{p\left(\omega \right)+\mu _{\textsl{QD}} R\left(\omega \right)}{r} k_{0}^{2} \exp \left(ik_{0} md\, sin\left(\theta \right)\right),
\end{equation}
\begin{equation} \label{EQ__22_} 
\begin{array}{ll} \displaystyle E_{total} & \displaystyle =
\frac{p\left(\omega \right)+\mu _{\textsl{QD}} R\left(\omega \right)}{r} k_{0}^{2} \sum _{m=0}^{N_{p} }\exp \left(ik_{0} md\, sin\left(\theta \right)\right) = 
 \\[\normalbaselineskip] & \displaystyle  = _{\left[sin\left(\theta \right)=0\right]}
\, \frac{p\left(\omega \right)+\mu _{\textsl{QD}} R\left(\omega \right)}{r} k_{0}^{2} N_{p}.
 \end{array}
\end{equation} 
In this paper, it was assumed that \textbf{$sin\left(\theta \right)=0$} (perpendicular transmission). All intensities were calculated at the distance of  $r_{0} =1\, {\rm m}$ from the chain of  $N_{p} =400$ particles.  
The scattered field intensity is $S_{total} =\frac{c}{8\pi } \left|E_{total} \right|^{2} $. This scattered field is pretty easy to measure in case of the spaser chain, because of the pump and scattered light wavelengths are spectrally separated. From the other side, in the case of spaser chain there is no physically justified normalizing field and it makes sense to present the results in absolute values. It is also useful to introduce separately the field intensity generated by \textsl{NP} $S_{\textsl{NP}} =\frac{c}{8\pi } \left|\frac{p\left(\omega \right)}{r} k_{0}^{2} N_{p} \right|^{2} $, the field intensity generated by \textsl{QD} $S_{\textsl{QD}} =\frac{c}{8\pi } \left|\frac{\mu _{\textsl{QD}} R\left(\omega \right)}{r} k_{0}^{2} N_{p} \right|^{2} $, and characterize the far filed by the relation $a_{S,\textsl{QD}} =S_{\textsl{QD}} /S_{total} $, which indicates part of the total field generated by the \textsl{QD}s. 

It is worth noting that the amplitude measurements (measurements of $S_{total} $) is not the only option. In case of $A_{ext} \left(\omega \right)\ne 0$, the much higher external field intensity makes it difficult to separate the significantly weaker scattered field $S_{total} $. In this case, interference between both fields (heterodyne detection) could provide an alternative measurement technique. The detector measures (in addition to the intensities of both transmitted and scattered fields) the interference part according to the known expression $\frac{c}{8\pi } \left|A_{ext} \left(\omega \right)+E_{total} \left(\omega \right)\right|^{2} =S_{ext} +S_{total} +\frac{c}{8\pi } \left(A_{ext}^{*} \left(\omega \right)E_{total} \left(\omega \right)+A_{ext} \left(\omega \right)E_{total}^{*} \left(\omega \right)\right)$. Lock-in technique allows us to separate $S_{ext} +S_{total} $ from the rest of the sum and measure the interference part only; here it is convenient to introduce the normalized intensity $a_{S} \left(\omega \right)$and beating part $a_{Int} \left(\omega \right)$ on external field intensity: 
\begin{equation} \label{EQ__23_} 
\begin{array}{ll}  \displaystyle
a_{S} \left(\omega \right)&= S_{total} \left(\omega \right)/S_{ext} \left(\omega \right), 
 \\[0.2\normalbaselineskip] \displaystyle
a_{Int} \left(\omega \right) &= 2\textsl{Re} \left[
A_{ext}^{*} \left(\omega \right)E_{total} \left(\omega \right) \right] / \left|A_{ext} \left(\omega \right)\right|^{2}= \\ &= 2\sqrt{a_{S}} \cos \left(\Delta \varphi \right). \end{array} 
\end{equation} 
Here $\Delta \varphi $ is the phase difference between the external and scattered (total) fields. This characteristic is shown to be more robust to the noise and thus far provides several orders of magnitude higher sensitivity in sensing applications \cite{Kravetz13}. 

\section{\label{sec:ChSpas}Chain of coupled spasers ($A_{ext} \left(\omega \right)=0$)}

The role of the saturation type nonlinearity of the \textsl{QD} coupled with \textsl{NP}, and its influence on the NBR is the main focus of this paper. In the case of lasing (spaser in generation operation mode) the subdivision of a ``linear'' and ``nonlinear'' dynamics does not make sense: the stationary state is stipulated by a nonlinearity (saturation) and hence is essentially nonlinear. 

In the case of an external field and low (below threshold) pumps the system exhibits its nonlinear properties only for particular set of parameters (sufficiently high external field intensities and/or significant pump levels).   

Consider solution of \eqref{EQ__17_} for the first spatial mode i.e. $k_{G} =0$, which in turn means equal phases and amplitude for all \textsl{NP}-\textsl{QD} in the chain. The system is in fact 1D array of the coupled spasers, and consequently exhibits typical laser properties, namely the system has a threshold and eigen generation frequency $\omega _{G}$. In the frame of this model the ``dark modes'' have the same threshold as the ``bright'' ones, in spite of the different levels of the emitted radiation. It is important to note either, that the spectrum bandwidth of the generation (whatever system -- single or coupled resonators in 1D array- is considered) is just delta function in the frame of this model; otherwise the stochastic fluctuations which result in the nonzero spectrum bandwidth have to be incorporated into the model \cite{Chirkin11}. 

\begin{figure}[b]
\includegraphics[width=0.4\textwidth]{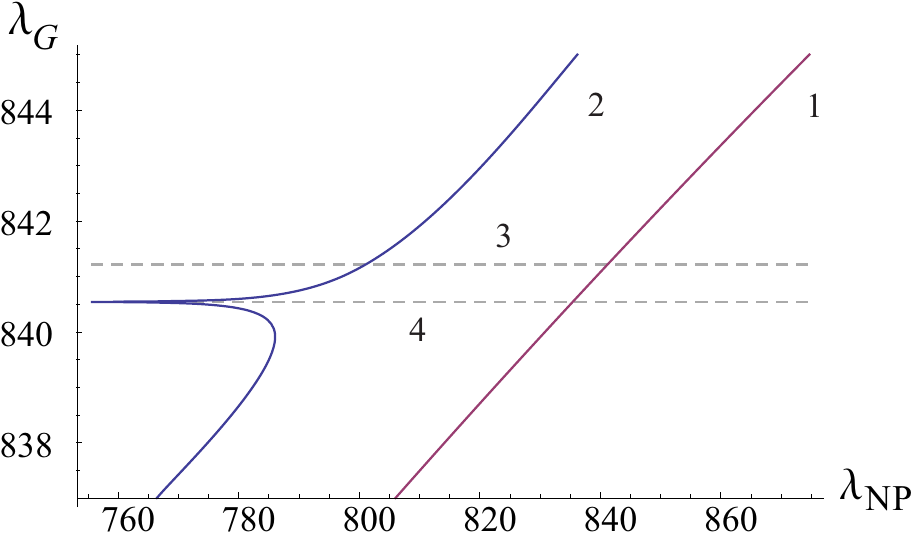} 
\caption{\label{fig4} Eigen generation wavelength $\lambda _{G,single} $ (curve 1) and eigen chain generation frequency $\lambda _{G} $ (curve 2) as a function of the eigen \textsl{NP} generation wavelength $\lambda _{\textsl{NP}} $.  Horizontal dashed lines mark the center of \textsl{QD} resonance line, $\lambda _{\textsl{QD}} =841.2\, {\rm nm}$ (curve 3), and the center of narrow band resonance (see below), $840.5\, {\rm nm}$ (curve 4).
} \end{figure}

In the case of absence of the external field system \eqref{EQ__19_} is reduced to:
\begin{equation} \label{EQ__24_} \left\{
\begin{array}{l} { \displaystyle
N_{st} \left(\omega _{G} \right)=N_{0} +\frac{\mu _{\textsl{QD}} }{2\hbar \gamma _{1} r_{\textsl{QD}}^{3} } \textsl{Im}\left[p\left(\omega _{G} \right)R^{*} \left(\omega _{G} \right)\right],
} \\[0.5\normalbaselineskip] { \displaystyle
R\left(\omega _{G} \right)=\frac{i\mu _{\textsl{QD}} }{\hbar \left(\gamma _{2} -i\delta \omega _{\textsl{QD},G} \right)} \frac{p\left(\omega _{G} \right)}{r_{\textsl{QD}}^{3} } N_{st} \left(\omega _{G} \right),
} \\[\normalbaselineskip] { \displaystyle
p\left(\omega _{G} \right)=-\alpha _{\textsl{NP}} \left(\omega _{G} \right)\frac{\mu _{\textsl{QD}} R\left(\omega _{G} \right)}{r_{\textsl{QD}}^{3} } .
} 
 \end{array} \right.
\end{equation}
which is the typical system describing the generation at eigen frequency $\omega _{G} $ which is also supposed to be resulted from \eqref{EQ__24_}. Combining second and third equations in \eqref{EQ__24_}, we obtain an equation for the stationary inversion $N_{st} $ and respective eigen mode frequency $\omega _{G} $: 
\begin{equation} \label{EQ__25_} \left\{
\begin{array}{l} { \displaystyle
N_{st} \left(\omega _{G} \right)=-\frac{\hbar \varepsilon _{h} r_{\textsl{QD}}^{6} }{\tilde{\mu }_{\textsl{QD}}^{2} } \text{Im}\left[\frac{\gamma _{2} -i\delta \omega _{\textsl{QD},G} }{\alpha _{\textsl{NP}} \left(\omega _{G} \right)} \right],
 } \\[0.5\normalbaselineskip] { \displaystyle
 \text{Re}\left[\frac{\gamma _{2} -i\delta \omega _{\textsl{QD},G} }{\alpha _{\textsl{NP}} \left(\omega _{G} \right)} \right]=0\, \, \, \Leftrightarrow 
 } \\[\normalbaselineskip] { \displaystyle
 \frac{\delta \omega _{\textsl{QD},G} }{\gamma _{2} } +\frac{\delta \omega _{\textsl{NP},G} +\chi _{\textsl{NP}} \text{Re}S_{1} \left(\omega _{G} \right)}{\gamma _{\textsl{NP}} +\chi _{\textsl{NP}} \text{Im}S_{1} \left(\omega _{G} \right)} =0\, \, \to \, \, \omega _{G} ,
 } \\[0.5\normalbaselineskip] {
 \delta \omega _{\textsl{QD},G} =\omega _{G} -\omega _{21} ,
 \, \delta \omega _{\textsl{NP},G} =\omega _{G} -\omega _{\textsl{NP}}.} \end{array} \right.
\end{equation}
\begin{widetext}
With known stationary inversion and eigen frequency, the stationary intensities $\left|R\left(\omega _{G} \right)\right|^{2} $and $\left|p\left(\omega _{G} \right)\right|^{2} $ are:
\begin{equation} \label{EQ__26a_}  
\begin{array}{l} { \displaystyle
\left|R\left(\omega _{G} \right)\right|^{2} =\frac{\gamma _{1} }{\gamma _{2} } N_{st} \left(\omega _{G} \right)\left[N_{0} +\frac{\hbar \varepsilon _{h} r_{\textsl{QD}}^{6} }{\tilde{\mu }_{\textsl{QD}}^{2} } \textsl{Im}\left(\frac{\gamma _{2} -i\delta \omega _{\textsl{QD},G} }{\alpha _{\textsl{NP}} \left(\omega _{G} \right)} \right)\right],
} \\[0.5\normalbaselineskip] { \displaystyle
\left|p\left(\omega _{G} \right)\right|^{2} =\frac{\hbar \gamma _{1} }{\varepsilon _{h} } \frac{\left|\alpha _{\textsl{NP}} \left(\omega _{G} \right)\right|^{2} }{\text{Im}\alpha _{\textsl{NP}} \left(\omega _{G} \right)} \left[N_{0} +\frac{\hbar \varepsilon _{h} r_{\textsl{QD}}^{6} }{\tilde{\mu }_{\textsl{QD}}^{2} } \text{Im}\left(\frac{\gamma _{2} -i\delta \omega _{\textsl{QD},G} }{\alpha _{\textsl{NP}} \left(\omega _{G} \right)} \right)\right],}
\end{array} 
\end{equation}
\noindent or substituting $\alpha _{\textsl{NP}} \left(\omega _{G} \right)$ from \eqref{EQ__24_}:
\begin{equation} \label{EQ__26b_}  
\begin{array}{l} { \displaystyle
\left|R\left(\omega _{G} \right)\right|^{2} =\frac{\gamma _{1} }{\gamma _{2} } N_{st} \left(\omega _{G} \right)\left[N_{0} -\frac{\hbar \varepsilon _{h} r_{\textsl{QD}}^{6} }{\tilde{\mu }_{\textsl{QD}}^{2} \chi _{\textsl{NP}} \gamma _{2} } \left(\gamma _{\textsl{NP}} +\chi _{\textsl{NP}} \text{Im}S_{1} \left(\omega _{G} \right)\right)\left(\gamma _{2}^{2} +\delta \omega _{\textsl{QD},G}^{2} \right)\right],} \\[0.5\normalbaselineskip] { \displaystyle
\left|p\left(\omega _{G} \right)\right|^{2} =\frac{\hbar \gamma _{1} }{\varepsilon _{h} } \frac{\chi _{\textsl{NP}} }{\gamma _{\textsl{NP}} +\chi _{\textsl{NP}} ImS_{1} \left(\omega _{G} \right)} \left[N_{0} -\frac{\hbar \varepsilon _{h} r_{\textsl{QD}}^{6} }{\tilde{\mu }_{\textsl{QD}}^{2} \chi _{\textsl{NP}} \gamma _{2} } \left(\gamma _{\textsl{NP}} +\chi _{\textsl{NP}} \text{Im}S_{1} \left(\omega _{G} \right)\right)\left(\gamma _{2}^{2} +\delta \omega _{\textsl{QD},G}^{2} \right)\right].} \end{array} 
\end{equation}
Pump threshold is determined by the evident requirement of the positive values of both intensities: 
\begin{equation} \label{EQ__26c_} 
N_{th} =-\frac{\hbar \varepsilon _{h} r_{\textsl{QD}}^{6} }{\tilde{\mu }_{\textsl{QD}}^{2} } \text{Im}\left[\frac{\gamma _{2} -i\delta \omega _{\textsl{QD},G} }{\alpha _{\textsl{NP}} \left(\omega _{G} \right)} \right]=\frac{\hbar \varepsilon _{h} r_{\textsl{QD}}^{6} }{\tilde{\mu }_{\textsl{QD}}^{2} \chi _{\textsl{NP}} \gamma _{2} } \left(\gamma _{\textsl{NP}} +\chi _{\textsl{NP}} \text{Im}S_{1} \left(\omega _{G} \right)\right)\left(\gamma _{2}^{2} +\delta \omega _{\textsl{QD},G}^{2} \right),  
\end{equation} 
\end{widetext}
which coincides with the threshold inversion \eqref{EQ__25_}. In particular, for a single \textsl{NP}-\textsl{QD} ($S_{1} \left(\omega _{G} \right)=0$ and $\alpha _{\textsl{NP}} \left(\omega _{G} \right)=i\chi _{\textsl{NP}} /\left(\gamma _{\textsl{NP}} -i\delta \omega _{\textsl{NP},G} \right)$), the stationary inversion, eigen frequencies, and stationary intensities are reduced to the known expressions:
\begin{equation} \label{EQ__27_}  
\begin{array}{l} { N_{th,single} =
\frac{\hbar \varepsilon _{h} r_{\textsl{QD}}^{6} \gamma _{2} \gamma _{\textsl{NP}} }{\tilde{\mu }_{\textsl{QD}}^{2} \chi _{\textsl{NP}} } \left[1+\left(\frac{\omega _{21} -\omega _{\textsl{NP}} }{\gamma _{2} +\gamma _{\textsl{NP}} } \right)^{2} \right],
} \\[0.5\normalbaselineskip] { \omega _{G,single} =
 \frac{\gamma _{2} \omega _{\textsl{NP}} +\gamma _{\textsl{NP}} \omega _{21} }{\gamma _{2} +\gamma _{\textsl{NP}} }, } \\[0.5\normalbaselineskip] { 
\left|R_{single} \right|^{2} =\frac{\gamma _{1} }{\gamma _{2} } N_{st,single} \left[N_{0} - N_{th,single}\right], } \\[0.5\normalbaselineskip] { 
\left|p_{single} \right|^{2} =\frac{\hbar \gamma _{1} }{\varepsilon _{h} } \frac{\chi _{\textsl{NP}} }{\gamma _{\textsl{NP}} } \left[ N_{0} - N_{th,single}\right].
} \end{array} 
\end{equation} 
Results of the spaser chain are presented in figures. First, the eigen generation wavelength dependence on the \textsl{NP} eigen wavelength $\lambda _{G} =\lambda _{G} \left(\lambda _{\textsl{NP}} \right)$ is shown in Fig.~\ref{fig4} for a single \textsl{NP}-\textsl{QD} (see (\eqref{EQ__27_})) and for the chain of spasers (see (\eqref{EQ__25_})).

This figure gives an initial comparative data about sensitivity of the single and a chain of the spasers to the environmental change, e.g. sensor operation. The refractive index changes causes shift of the eigen wavelength \textbf{$\lambda _{\textsl{NP}} $}which in turn leads to the observed shift of causes eigen resonance shift of $\lambda _{G} $. There is a special point around \textbf{$\lambda _{\textsl{NP}} =783\, {\rm nm}$} where even very small changes in \textbf{$\lambda _{\textsl{NP}} $} causes significant changes in $\lambda _{G} $, so that $\frac{\partial \lambda _{G} }{\partial \lambda _{\textsl{NP}} } \to \infty $. This effect is caused by a coupling in the chain and demonstrates potential of the considered system for the sensitivity improvements. 

Next set of pictures demonstrate threshold as a function of the \textsl{NP} wavelength at the fixed resonance \textsl{QD} wavelength for the single (see (\eqref{EQ__27_})) and a chain of the spasers (see (\eqref{EQ__26c_})) at environmental dielectric constant$\varepsilon _{h} =2.07$.
\begin{figure}[b]
\includegraphics[width=0.4\textwidth]{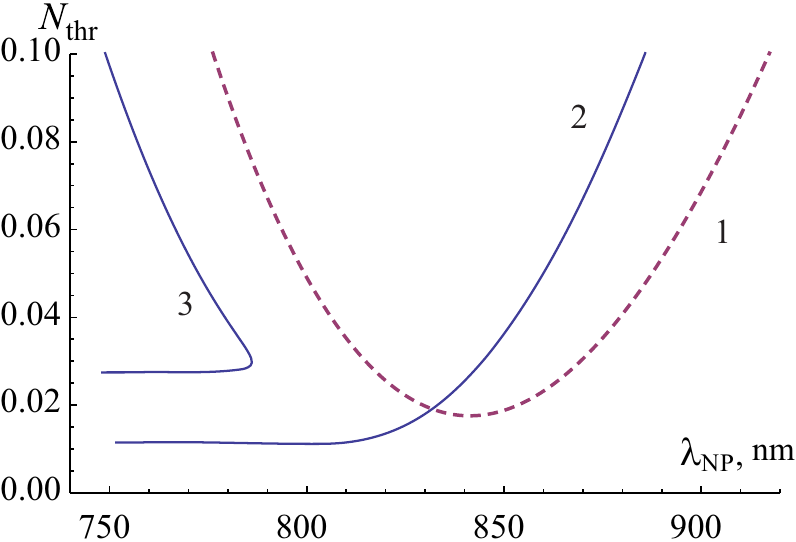} 
\caption{\label{fig5} Generation thresholds for the single (dashed curve 1) and chain of the spasers (curve 2, 3) as a function of the eigen \textsl{NP} generation wavelength  $\lambda _{\textsl{NP}} $. Hysteresis type curve 3 corresponds to the multi branch solution in Fig.~\ref{fig4} (see curve 2 in Fig.\ref{fig4}).}
\end{figure}
The resonant frequency for the spaser chain turns out to be shifted in blue region by several tens of nanometers. The pump threshold (in our notation) appears to be about $N_{0} =0,01$. This pretty low value is a consequence of the relatively high dipole moment for the \textsl{QD} accepted here for the calculation $\mu _{\textsl{QD}} =20\, Debai$, which definitely overestimates experimental values. The high value of the dipole moments leads in turn to lower saturation power and finally to the more pronounced nonlinearity. Nevertheless, we keep this numerical value in order to highlight the main physical effects of the nonlinear caused bistability and to demonstrate its potential for the sensor applications. 

\begin{figure}[b]
\includegraphics[width=0.4\textwidth]{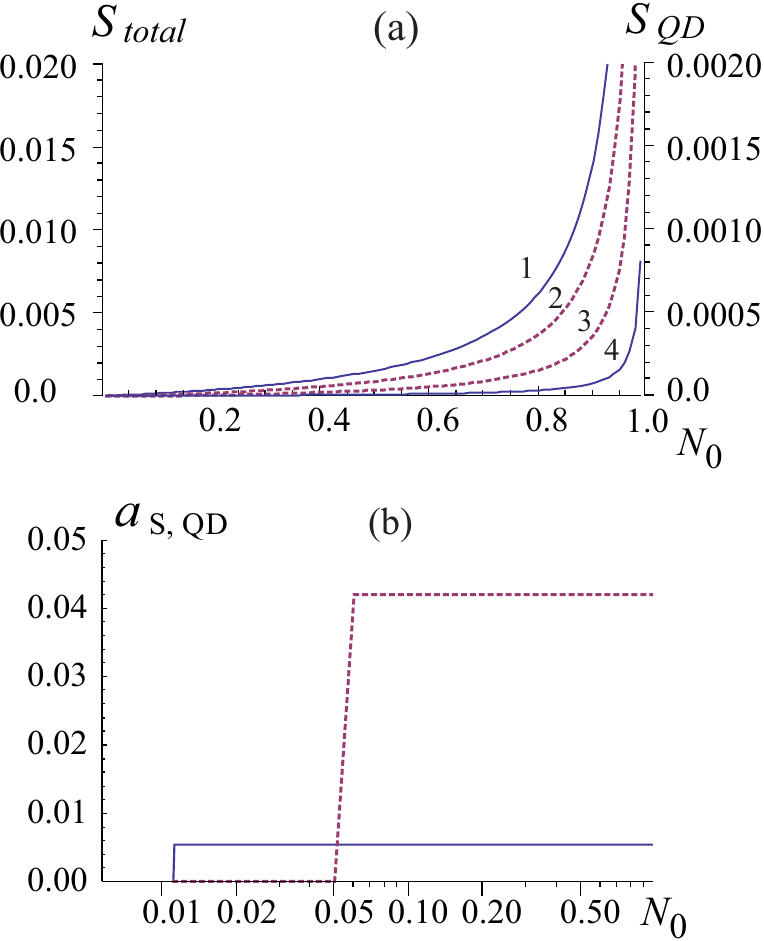} 
\caption{\label{fig6} (a) The stationary intensities, $S_{total} $ (curves 1 -- chain of spasers, 2 -- single spaser), $S_{\textsl{QD}} $ (curves 3 - single spaser, 4 -- chain of spasers), in $nW/cm^{2} $, and (b) relative stationary intensity $a_{S,\textsl{QD}} =S_{\textsl{QD}} /S_{total} $ for single (dashed curve) and chain of spasers (solid curve).}
\end{figure}

The stationary \textsl{QD} generated intensity $S_{\textsl{QD}} $, the total generated intensity $S_{total} $, and the relative intensity of \textsl{QD} $a_{S,\textsl{QD}} \left(\omega \right)$ in the far field zone are presented as functions of the pump $N_{0} $ in Fig.~\ref{fig6} for the single and chain of spasers. 

\section{\label{sec:ChNan}Chain of coupled nanoresonators driven by an external field ($A_{ext} \left(\omega \right)\ne 0$)}

\begin{figure}[t]
\includegraphics[width=0.4\textwidth]{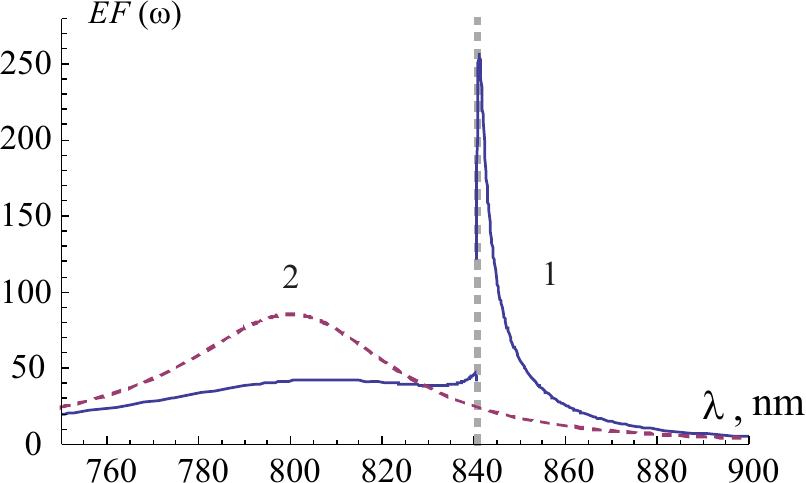} 
\caption{\label{fig7} Spectra of the enhancement factor $EF\left(\omega \right)=\left|1-\frac{\alpha _{\textsl{NP}} }{r_{\textsl{QD}}^{3} } \right|$ of the chain of spasers (curve 1) and of the single spaser (curve 2) for zero pump $N_{0} =-1$.  Vertical gray dashed line indicates position of the \textsl{QD} resonance.}
\end{figure}

\begin{figure}[t]
\includegraphics[width=0.4\textwidth]{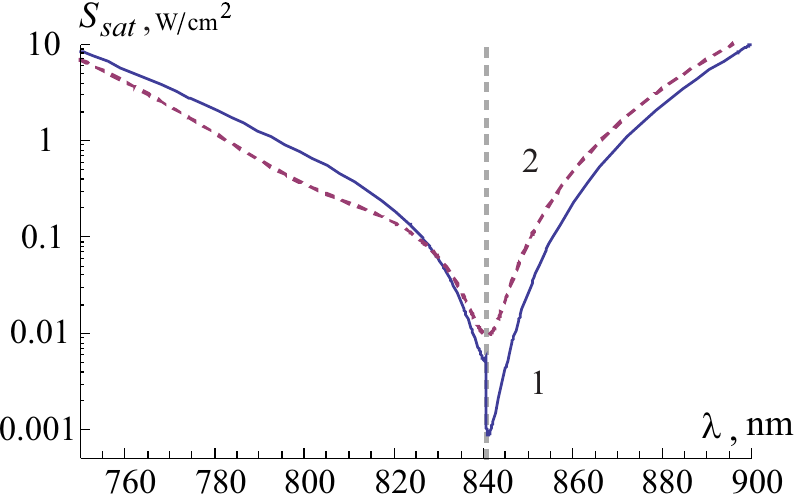} 
\caption{\label{fig8} Spectra of the saturation intensity $S_{sat} =\frac{c}{8\pi } \left|A_{sat} \left(\omega \right)\right|^{2} $ of the chain of spasers (curve 1) and of the single spaser (curve 2) for zero pump $N_{0} =-1$. Vertical gray dashed line indicates position of the \textsl{QD} resonance.}
\end{figure}

It is worth noting, that the nonlinearity is crucial for the appearance of the considered here resonances also in the case of the low (below threshold) levels of pump. The coupling between the \textsl{NP}s, described and quantified by the sums $S_{1} \left(\omega \right)$in \eqref{EQ__19_}, enhances the nonlinear response of the chain. In order to evaluate this effect, let us consider system \eqref{EQ__19_} and assume no action from the \textsl{QD} on the \textsl{NP}s (inverse action from the \textsl{NP} to \textsl{QD} is kept). In the case of the driving external filed it is justified, because of the external field intensities are typically much higher than ones generated by the \textsl{QD}s:    
\begin{equation} \label{EQ__28_} 
\left\{\begin{array}{l} {
 N\left(\omega \right)=N_{0} -\frac{\tilde{\mu }_{\textsl{QD}} }{2\hbar \gamma _{1} } 
 \text{Im}\left[ \left(A_{ext} \left(\omega \right)-
 \frac{p\left(\omega \right)}{r_{\textsl{QD}}^{3} } \right)R^{*} \left(\omega \right) \right],
  } \\[0.5\normalbaselineskip] { 
  R\left(\omega \right)=-\frac{i\tilde{\mu }_{\textsl{QD}} }{\hbar \left(\gamma _{2} -i\delta \omega _{\textsl{QD}} \right)} \left(A_{ext} \left(\omega \right)-
  \frac{p\left(\omega \right)}{r_{\textsl{QD}}^{3} } \right)N\left(\omega \right),
  } \\[0.5\normalbaselineskip] {
 p\left(\omega \right)=\alpha _{\textsl{NP}} \left(\omega \right)A_{ext} \left(\omega \right).} \end{array}\right.  
\end{equation} 
Substituting $p\left(\omega \right)$and$R\left(\omega \right)$into the first equation of \eqref{EQ__28_}, one can express $N\left(\omega \right)$ as a function of the external field intensity $\left|A_{ext} \left(\omega \right)\right|^{2} $, i.e. describe the saturation effect responsible for the nonlinear response:
\begin{equation} \label{EQ__29_} 
N\left(\omega \right)=\frac{N_{0} }{1+\frac{\tilde{\mu }_{\textsl{QD}}^{2} \left|A_{ext} \left(\omega \right)\right|^{2} }{\hbar ^{2} \gamma _{1} \gamma _{2} \left(1+\left(\frac{\delta \omega _{\textsl{QD}}^{} }{\gamma _{2}^{} } \right)^{2} \right)} \left|1-\frac{\alpha _{\textsl{NP}} \left(\omega \right)}{r_{\textsl{QD}}^{3} } \right|^{2}. }
\end{equation} 
It is clearly seen, that in addition to the usual saturation term $\frac{\tilde{\mu }_{\textsl{QD}}^{2} \left|A_{ext} \left(\omega \right)\right|^{2} }{\hbar ^{2} \gamma _{1} \gamma _{2} \left(1+\left(\frac{\delta \omega _{\textsl{QD}}^{} }{\gamma _{2}^{} } \right)^{2} \right)} $ there is one more term, which describes the nonlinearity enhancement effect due to the coupling in a chain$\left|1-\frac{\alpha _{\textsl{NP}} }{r_{\textsl{QD}}^{3} } \right|^{2} $ . It is convenient (and commonly accepted) to introduce the saturation intensity:
\begin{equation} \label{EQ__30_} 
\begin{array}{l} {
N\left(\omega \right)=\frac{N_{0} }{1+\frac{\left|A_{ext} \left(\omega \right)\right|^{2} }{\left|A_{sat} \left(\omega \right)\right|^{2} } }, } \\ {   S_{sat} \left(\omega \right)=\left|A_{sat} \left(\omega \right)\right|^{2} =\frac{\hbar ^{2} \gamma _{1} \gamma _{2} }{\tilde{\mu }_{\textsl{QD}}^{2} } \frac{\left(1+\left(\frac{\delta \omega _{\textsl{QD}}^{} }{\gamma _{2}^{} } \right)^{2} \right)}{\left|1-\frac{\alpha _{\textsl{NP}} }{r_{\textsl{QD}}^{3} } \right|^{2}.}  }
\end{array}
\end{equation} 
The lower saturation intensity $S_{sat} \left(\omega \right)$, the stronger nonlinear response of the chain. The saturation intensity $S_{sat} \left(\omega \right)$ decreases as the expression containing susceptibility $\left|1-\frac{\alpha _{\textsl{NP}} }{r_{\textsl{QD}}^{3} } \right|$  increases. This in turn takes place at  $S_{1} \left(\omega \right)\to \frac{\gamma _{\textsl{NP}} -i\delta \omega _{\textsl{NP}} }{i\chi _{\textsl{NP}} } $, see \eqref{EQ__19_}. In order to visualize this effect, both enhancement factor $EF\left(\omega \right)=\left|1-\frac{\alpha _{\textsl{NP}} }{r_{\textsl{QD}}^{3} } \right|$ and $\left|A_{sat} \left(\omega \right)\right|^{2} $ are plotted in Fig.~\ref{fig7}, \ref{fig8} for zero pump $N_{0} =-1$. Dependency of the saturation intensity \eqref{EQ__30_} from the pump level is defined by the energy relaxation rates $\gamma _{1} =2\tilde{\gamma }_{1} /\left(1-N_{0} \right)$, e. g. $S_{sat} \left(\omega \right)$ increases with growth of $N_{0} $. It means that the pump works ``against'' the nonlinearity enhancement and has to be avoided.  

The maximum of the enhancement factor $EF$ corresponds to the minimum of the saturation intensity $\left|A_{sat} \left(\omega \right)\right|^{2} $ and hence the respective wavelengths are primary candidates for ``working points'' of the chain: at these wavelengths the chain exhibits maximum nonlinear response. Minima of the curves in Fig.~\ref{fig8}, which coincide with center of \textsl{QD} line, present saturation intensity for the chain of spasers and single spaser respectively. The values of intensities differ by an order of magnitude, $S_{sat} =0.88\, \, {\rm mW/cm}^{{\rm 2}} $ for the chain and $S_{sat} =9.5\, {\rm mW/cm}^{{\rm 2}} $ for the single spaser.  

\begin{figure}[t]
\includegraphics[width=0.47\textwidth]{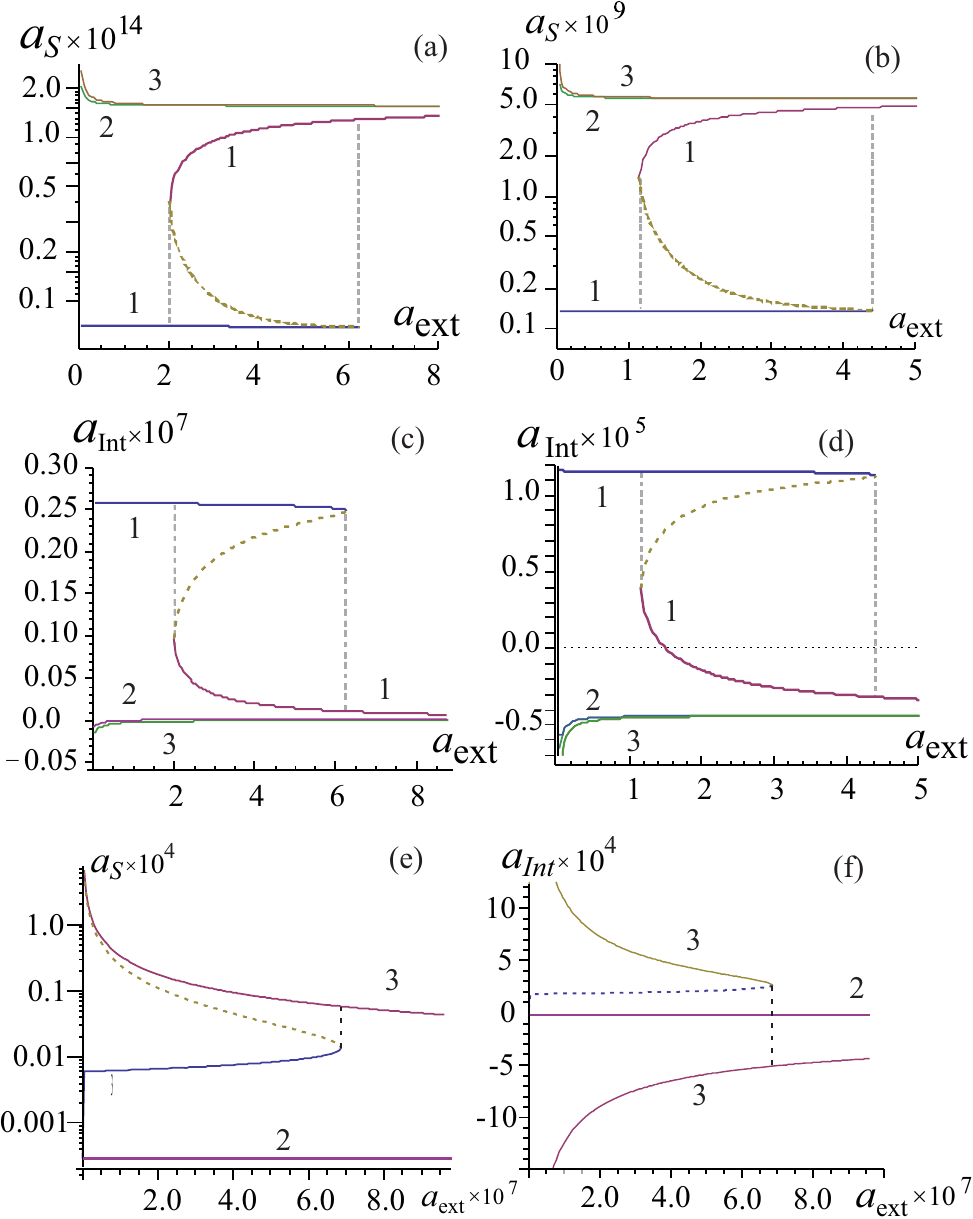} 
\caption{\label{fig9} Normalized field intensity $a_{S} $in far field zone for (a) single \textsl{NP}-\textsl{QD} and (b) chain of the \textsl{NP}-\textsl{QD} as a function of the external field intensity$a_{ext} $; interference signal $a_{Int} $ in far field zone for (c) single \textsl{NP}-\textsl{QD} and (d) chain of the \textsl{NP}-\textsl{QD} as a function of the external field intensity $a_{ext} $ at the pump levels $N_{0} =-1$ (hysteretic curve 1), $N_{0} =0,01$ (curve 2, hysteresis is absent), $N_{0} =0,02$ (curve 3, hysteresis too small);  (e,f) - $a_{S} $ and $a_{Int} $ for the chain of the \textsl{NP}-\textsl{QD} for two values of power below (curve 2) and above (curve 3) threshold. The wavelength of the external field coincides with the resonance wavelength of the \textsl{QD} and single \textsl{NP} (a), (c), or chain of \textsl{NP} (b), (d) , (e,f).}
\end{figure}

\begin{figure}[t]
\includegraphics[width=0.47\textwidth]{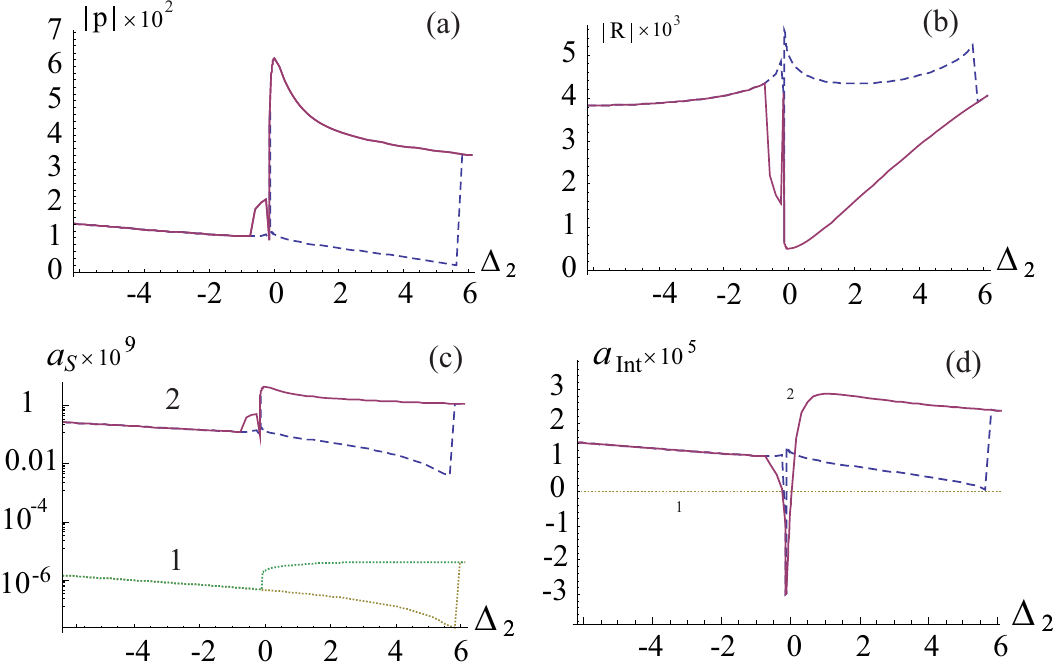} 
\caption{\label{fig10} Spectrum dependence of a single and a chain of the \textsl{NP}-\textsl{QD} in case of zero pump $N_{0} =-1$ in the middle of the hysteresis interval of the external field intensity$S_{ext} =0.77\, {\rm W/cm}^{{\rm 2}} $: (a) \textsl{NP} dipole moment amplitude $\left|p\right|$ (in units of $\mu _{\textsl{QD}} $),  (b) \textsl{QD} dipole moment amplitude $\left|R\right|$ all \textsl{QD}s in the chain, (c) normalized field intensity $a_{S} $ and (d) interference signal $a_{int} $\eqref{EQ__23_}. Dashed blue parts of the curves 2 correspond to lower branch of hysteresis. Dotted curve 1 in (b), (c) corresponds to a single \textsl{NP}-\textsl{QD} and solid curve 2 -- to the chain of \textsl{NP}-\textsl{QD}.}
\end{figure}

\begin{figure}[t]
\includegraphics[width=0.47\textwidth]{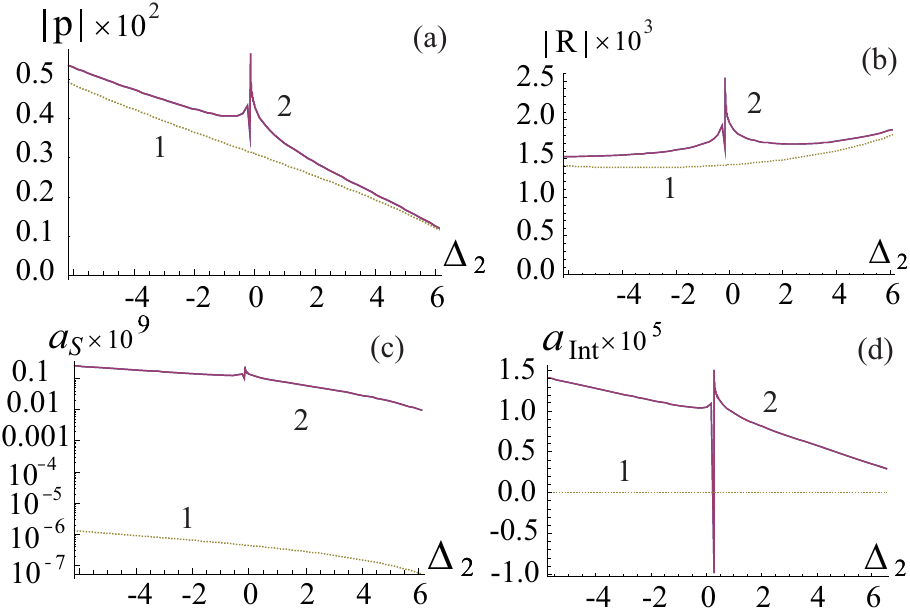} 
\caption{\label{fig11} Spectrum dependence of a single and a chain of the \textsl{NP}-\textsl{QD} in case of zero pump $N_{0} =-1$ and at the lower external field intensity $S_{ext} =0.117{\rm W/cm}^{{\rm 2}} $: (a) \textsl{NP} dipole moment amplitude $\left|p\right|$ (in units of $\mu _{\textsl{QD}} $, (b) \textsl{QD} dipole moment amplitude $\left|R\right|$ (c) normalized field intensity  $a_{s} $ \eqref{EQ__23_}, and (d) signal of the interference measurements $a_{int} $ \eqref{EQ__23_}. Dotted curve 1 corresponds to a single \textsl{NP}-\textsl{QD} and solid curve 2 -- to the chain of \textsl{NP}-\textsl{QD}.}
\end{figure}

\begin{figure}[t]
\includegraphics[width=0.47\textwidth]{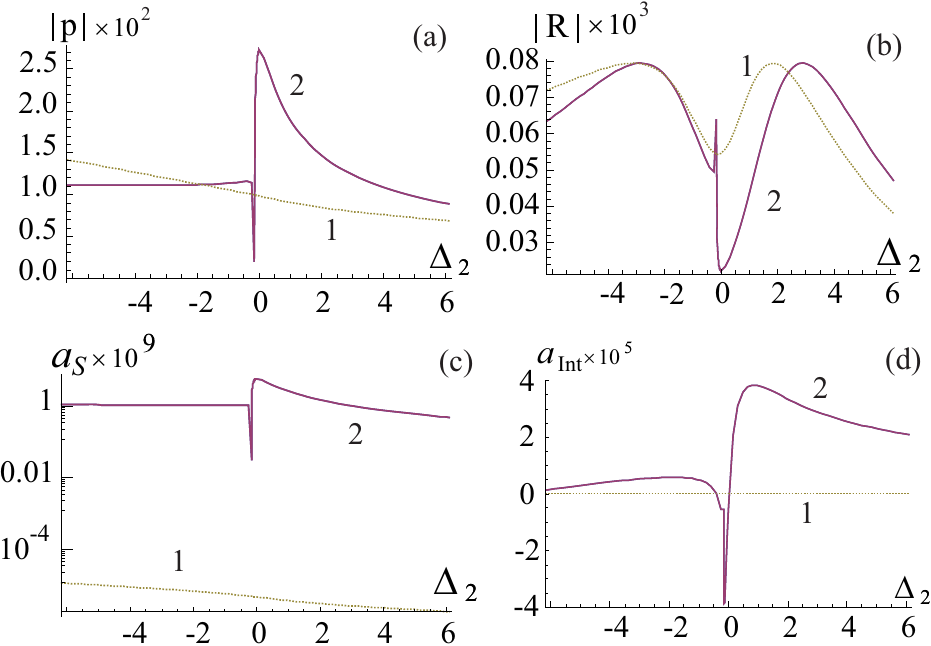} 
\caption{\label{fig12} Spectrum dependence of a single and a chain of the \textsl{NP}-\textsl{QD} in case of pump near threshold$N_{0} =0,01$ at the lower external field intensity $S_{ext} =0.117{\rm W/cm}^{{\rm 2}} $: (a) \textsl{NP} dipole moment amplitude $\left|p\right|$ (in units of $\mu _{\textsl{QD}} $), (b) \textsl{QD} dipole moment amplitude $\left|R\right|$, (c) normalized field intensity  $a_{s} $ and (d) interference signal $a_{int} $\eqref{EQ__23_}. Dotted curve 1 corresponds to a single \textsl{NP}-\textsl{QD} and solid curve 2 -- to the chain of \textsl{NP}-\textsl{QD}.}
\end{figure}

A distinguishable property of the response of the \textsl{NP}-\textsl{QD} (both single and chain) is the appearance of the hysteresis behavior. It is demonstrated in Fig.~\ref{fig9}, where the typical hysteresis curves of the normalized intensity $a_{S} $and interference signal $a_{Int} $are plotted for the case of resonance between the single \textsl{NP} and \textsl{QD}, $\omega _{21} =\omega _{\textsl{NP}} $ (maximum of the curve 4, on Fig.~\ref{fig7}), or resonance between chain of \textsl{NP} and \textsl{QD}, $\omega _{21} =\omega _{\textsl{NBR}} $ (maximum of the curve 2, on Fig.\ref{fig7}), at different pump levels. Natural constant for the normalization of the external field intensity is the value $\frac{\hbar ^{2} \gamma _{1} \gamma _{2} }{\tilde{\mu }_{\textsl{QD}}^{2} } $(see (29)), and the results hereafter are presented as a function of the normalized external field intensity $a_{ext} =\frac{\left|A_{ext} \left(\omega \right)\right|^{2} }{\left(\frac{\hbar ^{2} \gamma _{1} \gamma _{2} }{\tilde{\mu }_{\textsl{QD}}^{2} } \right)} $. In the case of pump level above threshold of the chain of \textsl{NP}-\textsl{QD} hysteresis behavior arises for essentially lower values of external field $a_{ext} <10^{-6} $ - see Fig.~\ref{fig9} (e), (f).
The dipole moments and respective field intensities as a function of the normalized frequency detuning $\Delta _{2} =\left(\omega _{21} -\omega \right)/\gamma _{2} $ are presented in Fig.~\ref{fig10} for the case of zero pump $N_{0} =-1$ and the external field intensity corresponding to the middle of the hysteresis range $S_{ext} =0.77\, {\rm W/cm}^{{\rm 2}} $. The hysteresis type spectrum curve can be achieved in the vicinity of the \textsl{QD} resonance, where the interaction between \textsl{NP} and \textsl{QD} reaches maximum. Influence of the coupling can be seen comparing the data in Fig.~\ref{fig10} for a single for a chain of the \textsl{NP}-\textsl{QD}.  The same set of data is presented in Fig.~\ref{fig11} for lower external field intensity $S_{ext} =0.117\, {\rm W/cm}^{{\rm 2}} $.  
Influence of the pump is presented in Fig.~\ref{fig12}, where the results for the same set of parameters are presented for the pump level near threshold $N_{0} =0.01$. 
From the presented above data one can conclude, that the narrowband resonances (which are of primary interest in this work) can be achieved if the edge of the hysteresis curve is shifted (by the external field intensity varying) to the resonance of the \textsl{QD} and \textsl{NP} $\Delta _{2} =\left(\omega _{21} -\omega \right)/\gamma _{2} \to 0$ (remind that the both resonances of the \textsl{QD} and chain of \textsl{NP} are supposed to coincide with each other. 

\begin{figure}[t]
\includegraphics[width=0.45\textwidth]{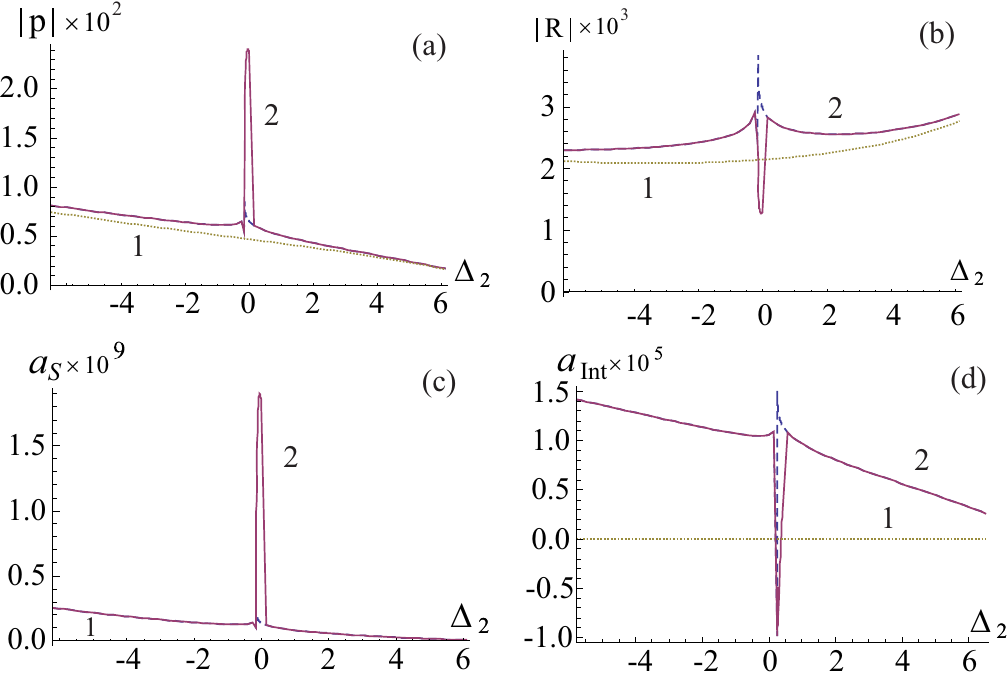} 
\caption{\label{fig13} Narrowband resonances at optimized external field intensities for a single and a chain of the \textsl{NP}-\textsl{QD} in case of zero pump $N_{0} =-1$ (curves 2, $S_{ext} =0.26\, 9{\rm W/cm}^{{\rm 2}} $) (a) \textsl{NP} dipole moment amplitude $\left|p\right|$ (in units of $\mu _{\textsl{QD}} $), (b) \textsl{QD} dipole moment amplitude $\left|R\right|$ (c) normalized scattered field intensity $a_{s} $ \eqref{EQ__23_}, and (d) signal of the interference measurements $a_{int} $\eqref{EQ__23_}. All data are presented for a single (dotted curves) and a chain of the \textsl{NP}-\textsl{QD} (solid curves).}
\end{figure}

\begin{figure}[t] 
\includegraphics[width=0.47\textwidth]{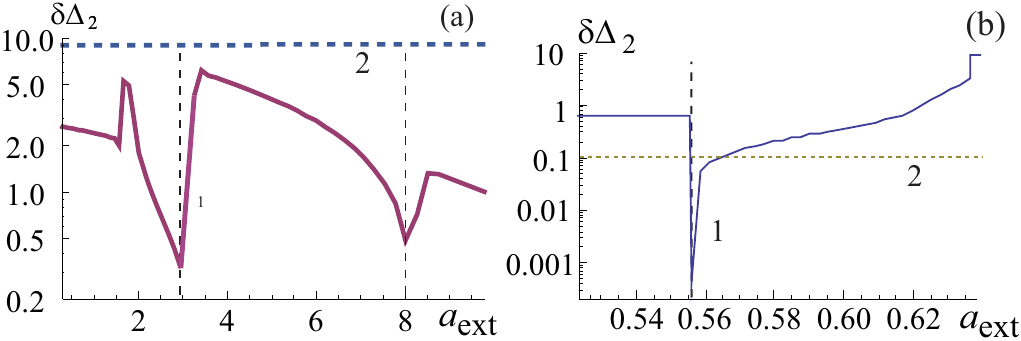} 
\caption{\label{fig14} Bandwidth of the found narrowband resonances of $a_{S} $ as a function of the normalized external field intensity: (a) for a single and (b) for a chain of the \textsl{NP}-\textsl{QD} in case of different pump levels (curve 1 - zero pump $N_{0} =-1$, curve 2 is  for the pump near threshold $N_{0} =0,01$). Dashed vertical markers denote the bounds of hysteresis region.}
\end{figure}

In Fig.~\ref{fig13} the obtained narrowband peaks at the optimum external field intensities are presented for a single and a chain of the \textsl{NP}-\textsl{QD}. 
The bandwidth of the found narrowband resonances depends significantly on the nonlinear properties of the considered system: the saturation nonlinearity of the \textsl{QD} causes the bandwidth $\Delta _{2} $ squeezing below approximately 0,1. In the case of low nonlinear response (or/and weak coupling between the \textsl{NP} and \textsl{QD}) the resonance bandwidth is in the range of $\Delta _{2} \sim 4\div 5$.  The results of the narrowband resonance bandwidth (for both amplitude $a_{S} $ and interference measurements $a_{Int} $) are presented in Fig.~\ref{fig14} as a function of the normalized external field intensity at the different pump levels. 

\section{\label{sec:Concl}Conclusion}

In conclusion, it is necessary to discuss the experimental procedure to visualize the found here narrowband resonances. The spectra shown in Figs.~\ref{fig10}-\ref{fig13} are not ones typically measured, for example, in transmission/reflection tests. The found spectra are response dependence on the wavelength of the driving external field causing the hysteresis behavior. In order to realize this spectrum experimentally, the hysteresis curves (e.g. Fig.~\ref{fig9}) have to be measured for different wavelengths of the external field. Nevertheless, there is a way to realize these spectra dependence in real time. In this case a high power driving field has to be accompanied by a wide band low intensity field used to measure spectrum. The high power field delivers the system (chain of \textsl{QD}-\textsl{NP}) to the upper hysteresis branch, and the spectrum has to be measured at the appropriate driving field intensity. The driving field is supposed to be pulsed as well with the typical time duration of about 100 ns, while the spectrum measurement has to be accomplished during about 10 ns at the appropriate time point of the driving pulse. Rigorously speaking, the theoretical data presenting in this paper do not consider the ``real-time'' spectrum measurement and have to be adopted for the respective procedure. This ``real-time'' spectrum measurement is supposed to be implemented for sensor applications and is a subject of a separated publication.   

The found in this the paper narrowband resonances are based on the nonlinear hysteresis phenomena and hence differ principally from any types of the linear resonances, e.g. presented in \cite{Kravets08} . In combination with the high sensitivity of the multi-stable systems, these narrow band resonances promise to be an alternative approach for different sensor applications.   

\begin{acknowledgments}
This work was partially financially supported by Government of Russian Federation, Grant 074-U01, and by RFBR Projects No. 12-02-00853-a (S.V.F).
\end{acknowledgments}



\end{document}